\documentclass[a4paper]{scrartcl}
\usepackage[utf8]{inputenc}
\usepackage{pgfplots}

\immediate\pdfobj stream attr{/N 3}  file{sRGBIEC1966-2.1.icm}
\pdfcatalog{%
/OutputIntents [ <<
/Type /OutputIntent
/S/GTS_PDFA1
/DestOutputProfile \the\pdflastobj\space 0 R
/OutputConditionIdentifier (sRGB IEC61966-2.1)
/Info(sRGB IEC61966-2.1)
>> ]
}

\usepackage{hyperxmp}
\usepackage[pdfa]{hyperref}

\hypersetup{
colorlinks,
linkcolor={red!50!black},
citecolor={green!50!black},
urlcolor={black},
pdftitle={Solving k-means on High-dimensional Big Data},
pdfauthor={Jan-Philipp W. Kappmeier; Daniel R. Schmidt; Melanie Schmidt},
pdfsubject={Preprint},
pdfkeywords={k-means clustering, data streams, SVD},
pdflang={en},
pdfpagelayout={OneColumn},
pdfstartpage={1},
bookmarksopen=true,
bookmarksopenlevel=0
}

\usepackage{tikz}

\usepackage{graphicx}

\usepackage[left=2cm,right=2cm,top=2cm,bottom=3cm]{geometry}

\usepackage{amsmath,amsfonts}
\usepackage{lscape}
\usepackage{booktabs}
\usepackage{longtable}
\usepackage{xspace}
\newcommand{\ie}{i.\nolinebreak[4]\hspace{0.125em}\nolinebreak[4]e.,\@\xspace}

\newcommand{\eg}{e.\nolinebreak[4]\hspace{0.125em}\nolinebreak[4]g.,\@\xspace}

\newcommand{\R}{\mathbb{R}}
\newcommand{\N}{\mathbb{N}}

\usetikzlibrary{chains,patterns,decorations.pathreplacing,shapes.geometric,calc}

\usepackage[scientific-notation=true]{siunitx}
\usepackage{authblk}

\author[1]{Jan-Philipp~W.~Kappmeier}
\author[2]{Daniel~R.~Schmidt}
\author[2]{Melanie~Schmidt}
\affil[1]{Technische Universität Berlin, Germany, \href{mailto:kappmeier@math.tu-berlin.de}{\texttt{kappmeier@math.tu-berlin.de}}   }
\affil[2]{Carnegie Mellon University, Pittsburgh PA, \texttt{\{\href{mailto:schmidtd@andrew.cmu.edu}{schmidtd},\href{mailto:mschmid1@andrew.cmu.edu}{mschmid1}\}@andrew.cmu.edu}}

\title{Solving $k$-means on High-dimensional Big Data} %

\begin{document}
\pagestyle{plain}
\maketitle              %

\begin{abstract}
In recent years, there have been major efforts to develop data stream algorithms that process inputs in one pass over the data with little memory requirement. For the $k$-means problem, this has led to the development of several $(1+\varepsilon)$-approximations (under the assumption that $k$ is a constant), but also to the design of algorithms that are extremely fast in practice and compute solutions of high accuracy. However, when not only the length of the stream is high but also the dimensionality of the input points, then current methods reach their limits.

We propose two algorithms, piecy and piecy-mr that are based on the recently developed data stream algorithm BICO that can process high dimensional data in one pass and output a solution of high quality. While piecy is suited for high dimensional data with a medium number of points, piecy-mr is meant for high dimensional data that comes in a very long stream.
We provide an extensive experimental study to evaluate piecy and piecy-mr that shows the strength of the new algorithms.
\end{abstract}

\section{Introduction}
Partitioning points into subsets (\emph{clusters}) with similar properties is an intuitive, old and central question.
\emph{Unsupervised} clustering aims at finding structure in data without the aid of class labels or an experts opinion.
It has many applications ranging from computer science applications like image segmentation or information retrieval to applications in other sciences like biology or physics where it is used on genome data and CERN experiments.
For an overview on the broad subject, see for example the survey by Jain~\cite{J10}.
The \emph{$k$-means problem} asks to cluster data such that the sum of the squared error is minimized.
It has been studied since the fifties~\cite{L57,S56} and optimizing it is likely \lq the most commonly used partitional clustering strategy\rq~\cite{JD88}.
It measures the quality of a partitioning of points from $\mathbb{R}^d$ based on the squared Euclidean distance function. Each cluster in the partitioning is represented by a center, and the objective function is the sum of the squared distances of all points to their respective center.

The popularity of the $k$-means problem is underlined by the fact that the most popular algorithm for it, Lloyd's algorithm, was named one of the ten most influential algorithms in the data mining community by the organizers of the IEEE International Conference on Data Mining (ICDM) in 2008, see Wu et.\ al.~\cite{WKQGYMMcLNLYZSHS08}.
Lloyd's algorithm~\cite{L82} (independently developed by Steinhaus~\cite{S56}) is a local search heuristic that iterates the following two steps. First, it obtains an initial solution consisting of $k$ centers, \eg by drawing $k$ centers uniformly at random from the input. Then, the following two steps are alternated: Assign every point to its closest center to obtain a partitioning into $k$ subsets, compute the centroid of each subset and replace the center by this centroid. Both steps can only decrease the cost. Assigning points to their closest center is optimal for the given centers, and for each subset, the centroid is the optimal center. Thus, the new solution is either cheaper or of equal cost. In the latter case, the algorithm has converged\footnote{Since there are finitely many partitionings, the algorithm eventually converges to a local optimum. It is also common to stop the algorithm after a predefined number of iterations, or when the decrease of the cost function is small.}.

The quality in terms of the sum of squared errors of the output of Lloyd's algorithm depends on the local optimum that is reached. Finding a good local optimum can be achieved by initializing the algorithm with a good initial solution. Arthur and Vassilvitskii~\cite{AV07} propose the \emph{$k$-means++} method as an improved version of Lloyd's algorithm. It chooses the initial solution randomly, but only the first center is chosen uniformly at random. The $i$th center is chosen by computing all points squared distances to their closest center and then chosing each point with a probability proportional to its cost as the next center. This way, it is likely that most optimal centers have a close center in the start solution. This initialization method produces centers which are an $\mathcal{O}(\log k)$-approximation in expectation, and experiments indicate that the local optimum found from this start solution is usually of high quality.

The $k$-means++ method therefore provides a great tool for solving the $k$-means problem in practice, with an (expected) worst-case guarantee, a very good practical performance and the advantage that it is very easy to implement. The theoretically best approximation algorithms for the $k$-means problem provide a constant factor approximation for the general case~\cite{JV01,KMNPSW04} and a $(1+\varepsilon)$-approximation (even in linear time) if $k$ and $\varepsilon$ are assumed to be constants~\cite{FL11}.

For big data, running Lloyd's algorithm or $k$-means++ is less viable. Asymptotically, the running time of both algorithms is $\mathcal{O}(ndk)$ if the number of iterations is bounded to a constant. This looks convincing since a straightforward implementation of finding the closest center for a point takes $\Theta(dk)$ time, so even evaluating a solution then has running time $\Theta(ndk)$. Additionally, the input size is already $\mathcal{O}(nd)$, so the running time is linear for constant values of $k$. However, both algorithms need random access to the data and iterate over it several times. As soon as the data does not fit into main memory, the algorithms do thus not scale very well. For example, $k$-means++ needed over seven hours to compute 50 centers for a 54-dimensional data set (\emph{Covertype}) with half a million points~\cite{AMRSLS12}.

A natural strategy to cope with this problem is to summarize the data before running the respective algorithm. A famous example for this is BIRCH~\cite{ZRL97}, a SIGMOD Test of Time Award winning algorithm that computes a summary by one pass over the input data and then clusters the points in the summary. BIRCH is very fast and thus enables the processing of large data sets. However, the quality in terms of the sum of squared errors can be low~\cite{AMRSLS12,FGSSS13}.

A more recent development is the design of fast data stream algorithms that are based on \emph{coresets}. A coreset $S$ of a point set $P$ is a weighted summary of $P$ that maintains a strong quality guarantee: For any choice $C$ of $k$ centers, the $k$-means costs of the clustering induced by $C$ on $S$ are within an  $(1+\varepsilon)$-factor of the $k$-means clustering that $C$ induces on $P$. Thus, executing any $k$-means algorithm on the coreset gives a good approximation of what the same algorithm would have produced on $P$. Coreset constructions are generally designed with a focus on strong theoretic bounds, but can be made viable in practice with slight heuristic changes.

StreamKM++~ is such an algorithm~\cite{AMRSLS12}. It computes a coreset in one pass over the data and then runs $k$-means++ on the coreset. The size of the coreset is polylogarithmic in the input sizes if the dimension of the data is constant. The total memory requirement is also polylogarithmic. Experiments show that the quality of the solutions is comparable to the $k$-means++ solutions (on the full data set) while the running time is a small fraction. For example, the above mentioned covertype is processed in ten minutes instead of seven hours, with a result of similar quality.

BICO is a recent algorithm that outperforms StreamKM++ on all data sets that are tested in~\cite{AMRSLS12,FGSSS13} and enables the processing of data sets with millions of points in less than an hour\footnote{One data set is \emph{BigCross}, containing three million points in 68 dimensions and is processed in under twenty minutes for $k\le 250$.}. The above mentioned test case needs 27 seconds instead of ten minutes for StreamKM++ and seven hours for $k$-means++, and larger instances show even higher acceleration. BICO is also based on a coreset construction, using a slight variation of an algorithm with a strong theoretical guarantee. The quality of the computed solutions in experiments is as good as that of StreamKM++. The source code of BICO is written in C++ and is available online.

For data sets with up to around $100$ dimensions, this is a pleasant state of affair. However, both the analysis of the running time and memory requirement of StreamKM++ and BICO assume that the dimension is a constant. At least for BICO, this is not a theoretically imposed restriction, but does indeed correspond to an unfavorable dependency on the dimension. The reason is that BICO covers the input data by spheres (in order to summarize all points in the same sphere by one point).
When the number of spheres is too large, a rebuilding step reduces it by merging certain spheres. Covering a set by spheres gets increasingly difficult as the dimension gets higher, which results in several rebuilding steps of BICO, and in a higher running time.

On the theoretical level, however, there are several results saying that it is possible to compute a coreset of a point set in one pass and with low memory requirements. For example, Feldman and Langberg~\cite{FL11} propose a one-pass algorithm that computes a coreset with storage size of $\mathcal{O}\left(k d \log^4 n \varepsilon^{-3}\log 1/\varepsilon\right)$. It is thus theoretically possible to compute coresets which scale well with the dimension, but there is no practical algorithm yet that achieves a high quality summary and can cope with very high dimensional, large data sets.

\subsection{Our Contribution}
We develop two new algorithms, \emph{piecy} and \emph{piecy-mr} that can deal with high-dimensional big data.
For that, we combine BICO with a dimensionality reduction. This reduction is done by projecting onto the best fit subspace (of a parameterized dimension) which can be computed by the \emph{singular value decomposition}~(SVD).
This is theoretically supported by recent results~\cite{CEMMP15,FSS13} that say that projecting onto the best fit subspace of dimension $\lceil k/\varepsilon\rceil$ and then solving the $k$-means problem gives a $(1+\varepsilon)$-approximation guarantee.
We find that $3k/2$ dimensions are often sufficient to give highly accurate results. This might be due to the spectrum of the data we used.

The next challenge is to intertwine the dimensionality reduction with the coreset computation in order to do both in one pass over the data. The first algorithm, \emph{piecy}, reads chunks (pieces) of the data and processes, reduces the dimensionality of each chunk and feeds the resulting points into BICO. The drawback of this approach is that the total dimensionality of the complete point set that is fed into BICO increases with the number of pieces. For large data sets and high input dimension, this approach will eventually run into the same trouble as BICO (but only for data sets that are larger and higher dimensional than those BICO can process).
In \emph{piecy-mr}, we resolve this potential limitation by adapting a technique called \emph{Merge-and-Reduce}~\cite{HPM04}. It is a method that shows that any coreset computation can be turned into a one-pass algorithm at the cost of additional polylogarithmic factors. We adapt it to take advantage of the fact that we use a coreset computation (BICO) which already \emph{is} a one-pass algorithm.

As intermediate steps of our work, we evaluate two implementations for the singular value decomposition, an implementation in Lapack++~\cite{LL10} and the implementation called redSVD~\cite{O11}. We compare their speed and quality. Furthermore, we extend the algorithm BICO to process weighted inputs (which is necessary for our piecy-mr approach).

\section{The algorithms}\label{sec:the-algorithms}
In the following, we describe the three algorithms that we tested: BICO and our two new algorithms, \emph{piecy} and \emph{piecy-mr}. For a point set $P$, we denote the centroid of $P$ by $\mu(P) := \sum_{x\in P} x / |P|$.

\subsection{BICO}
BICO uses a data structure based on \emph{clustering features}. A clustering feature of a point set $S$ consists of the number of points $|S|$, the sum of the points $\sum_{x \in S} x$ and the sum of the squared length of the points $\sum_{x \in S} x^tx$. By the well-known formula
\[
\sum_{x \in P} ||x-c||^2 = |P|\cdot ||\mu(P) - c||^2 + \sum_{x \in P}||x-\mu(P)||^2,
\]
which holds for every point set $P$, a clustering feature is enough to exactly compute the cost between a point set and \emph{one} center.
BICO uses spheres that cover the input data. The point set inside each sphere is represented by a clustering feature. When a point arrives, it can be added to a clustering feature in constant time. The challenge for BICO is to decide into which clustering feature a point shall be added in order to equally distribute the error and to keep the overall error small. This is achieved by managing the clustering features in a well organized tree. Finding an appropriate clustering feature to add a point dominates the insertion time of a point. It lies between $\Theta(1)$ and $\Theta(m)$ for each point, where $m$ is the coreset size. BICO includes several heuristics to speed up the identification process of a good clustering feature such that the running time is often closer to $\Theta(1)$ per point. How well these heuristics work depends on the dimension of the input point set.

Whenever the number of spheres (and thus clustering features) exceeds $m$, BICO performs a rebuilding step that merges some of the spheres and their features together. For high-dimensional data sets, this may occur more often unless the spheres become large enough. More rebuilding steps imply a higher running time.

\subsection{Piecy}

Our aim is to compute coresets for large high-dimensional data sets by using BICO and dimensionality reduction techniques, but in \emph{only one pass} over the data.
\emph{Piecy} pursues the idea of running only a single instantiation of BICO and subsequently feeding it with chunks of low dimensional points. Thus, piecy reads a piece of $p$ points, reduces its intrinsic dimension and inputs the resulting points into BICO.

\paragraph*{Choice of dimensionality reduction technique and number of dimensions.}
We use the projection to the best fit subspace of dimension $\ell$, where $\ell$ is a parameter to be optimized.
The best fit subspace can be computed by using the \emph{singular value decomposition}.
The theoretical background of this approach is that projecting to best fit subspaces yields a good approximation of the squared pairwise distances~\cite{CEMMP15,DFKVV04}. When projecting to $k$ dimensions, a $2$-approximation is guaranteed, while projecting to $\lceil k/\varepsilon\rceil$ guarantees a $(1+\varepsilon)$-approximation. Thus, we test values between $k$ and moderate multiples of $k$ to get a reasonable compromise between approximation factor and running time.

\paragraph*{Using SVD to project to the best fit subspace.}
When we say that we use \lq the\rq\ SVD, we mean the SVD of the matrix $A \in \mathbb{R}^{n\times d}$ where the input points are stored in the columns. The SVD of $A$ has the form $A = U D V^T$ for matrices $U \in \mathbb{R}^{n \times n}, D \in \mathbb{R}^{n \times d}, V\in \mathbb{R}^{d \times d}$, where $U$ and $V$ are unitary matrices and $D$ is a diagonal matrix. The matrix $V$ contains the right singular vectors of $A$.
The projection of (the points stored in) $A$ to the best fit subspace of dimension $\ell$ is the matrix $A_\ell = U D_\ell V^T$, where $D_\ell$ is obtained by replacing all but the first $\ell$ diagonal elements by zero. Notice that the resulting matrix still contains $d$-dimensional points, but their \emph{intrinsic} dimension is reduced to $\ell$. This still helps, since the $\ell$-dimensional point set is easier to cover for BICO.

\paragraph*{Computation of the SVD.}
Numerically stable computation of the singular value decomposition is a research field of its own. Basic methods that compute the \emph{full} SVD, e.g. $U$, $V$ and $D$, have a running time of $\Omega\left(n d \min\left(n,d\right)\right)$. This full SVD can be used by dropping the appropriate entries of $D$ to obtain a matrix $D_\ell$ and evaluating the matrix product $U D_\ell V^t$ to obtain the projection onto the best fit subspace of dimension $\ell$. However, a variety of more efficient algorithms have been developed for this specific task, which are known as algorithms for the \emph{truncated} SVD that computes a decomposition $A_\ell = U_\ell D_\ell V_\ell^t$ directly without computing the full SVD of $A$. Additionally, random variations are known that reduce the running time sufficiently at the cost of a small error.
Mahoney~\cite{M11} gives a very nice overview on different methods to compute the singular value decomposition, then continuing with a detailed view on randomized methods and also discussing practical aspects.
For this work, we use an implementation that is based on the randomized algorithm presented in~\cite{HMT11} that multiplies $A$ with a randomly drawn matrix to reduce the number of its columns before computing the SVD. The implementation is called \emph{redSVD}~\cite{O11}. In addition to reducing the number of columns, it also reduces the number of rows before computing the SVD. Below, we experimentally compare the performance of redSVD to the performance of the \emph{lapack++} implementation of the full SVD computation.

\emph{Parameters.} The authors of BICO propose using a coreset size of $200k$ for BICO, which we adopt. That given, there are two parameters to be chosen: The size of the pieces that are the input for one SVD, and the number of dimensions we project to. As we argued above, the latter should be at least $k$ and not more than a reasonable multiple of $k$.

\emph{Memory requirement.} At each point in time, we store at most one piece of the input, one SVD object and one BICO object. The memory requirement of BICO is proportional to the output size, \ie to $200k$.

\emph{Obtaining a solution.} Running \emph{piecy} computes a summary of the input points. In order to obtain an actuall solution for the $k$-means problem, we run $k$-means++\cite{AV07} on the summary.

\subsection{Piecy-MR}
Notice that each chunk of data that is processed by piecy adds (in the worst case) $m$ dimensions to the intrinsic dimension of the point set that is stored by the BICO instance, as long as the maximum dimension is reached. For large data sets, this is unfavorable.

\paragraph*{Helpful coreset properties.}
A convenient property of coresets helps here. Assume that $S_1$ and $S_2$ are coresets for points sets $P_1$ and $P_2$, \ie their weighted cost approximates the weighted cost of $P_1$ or $P_2$, respectively, for any possible solution, and up to an $\varepsilon$-fraction. Then the weighted cost of their union $S_1 \cup S_2$ approximates the cost of $P_1\cup P_2$ for any solution up to an $\varepsilon$-fraction as well. Furthermore, if we use a coreset construction to reduce $S_1 \cup S_2$ to a smaller set (since $|S_1 \cup S_2|$ will be larger than the size of one coreset), then we obtain a coreset for $P_1 \cup P_2$. The error gets larger but is bounded by a $(3\varepsilon)$-fraction of the cost of $P_1\cup P_2$ (which can be compensated by choosing a smaller $\varepsilon$ to begin with).

\paragraph*{The Merge-and-Reduce technique.}
Assume for a moment that our aim is solely to compute a coreset with no thoughts about the intrinsic dimension of the points, but given a coreset computation that needs random access to the data. Then an intuitive approach is to read chunks of the data, computing a coreset for each chunk and joining it with previous corsets, until the union becomes too large. Then we could reduce the union by another coreset construction. The problem with this approach is that the first chunk of the data will participate in all following reduce steps, making the error unnecessary high. The Merge-and-Reduce technique~\cite{BS80} (for clustering for example used in~\cite{AHPV04,HPM04}) organizes the merge and reduce steps in a binary tree such that each point takes part in at most $\mathcal{O}(\log n)$ reduce steps for a stream of $n$ points.

\begin{figure}
\centering
\includegraphics[width=0.95\textwidth]{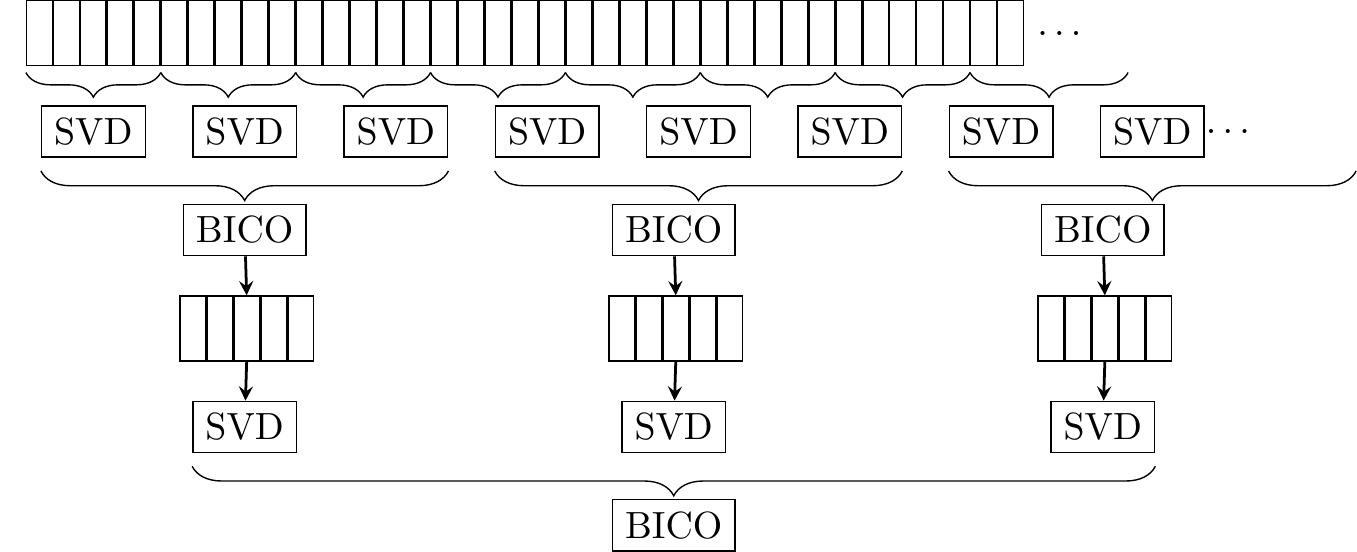}
\caption{\label{fig:piecy-mr-tree}The Merge-and-Reduce style tree built by piecy-mr
with an exemplary \emph{piece size} of $5$ and a \emph{number of pieces} of $3$.
Every chunk with \emph{piece size} many points is first fed into a singular value decomposition.
The result of the SVD contains the same number of points but has a smaller intrinsic dimension.
It is then fed into an instantiation of the BICO algorithm.
After \emph{number of pieces} many chunks, the BICO algorithm computes a coreset of size \emph{piece size}.
Thus, we can continue on the next layer. On each layer, the number of points is reduced by a
factor of \emph{number of pieces}. We continue to call the SVD on each layer to keep the intrinsic
dimension of the point set small.\label{fig:computationtreepiecymr}
}
\end{figure}

\paragraph*{Our computation tree.}
We have a different problem since the coreset construction that we use, BICO, does not require random access to the data. Instead, we wish to keep the dimension of the input data small. Assume we would consider this problem independently from the coreset computation, by just computing the SVD of chunks of the data and keeping the reduced points in memory (maybe performing a second pass over the data to compute the coreset). This is infeasible since the number of points is not reduced and hence we would store the complete data set (with a lower intrinsic dimension). Imagine even that at each point in time, an oracle could provide us with the best fit subspace of dimension $\ell$ of all points seen so far. We could still not easily use this information since the best fit subspace would change over time. So if we use one instance of BICO, and input each point into it, projected to the best fit subspace of all points seen so far, then we would still get a high intrinsic dimension for the points stored in BICO.

By also embedding BICO into the Merge-and-Reduce tree, we solve these problems. The first way of doing this would be to view the two steps of reducing the dimension and entering the points into BICO as one coreset computation, and just embed this into the Merge-and-Reduce technique. However, this has the drawback that we perform the same number of dimensionality reductions as we use BICO for reducing sets to smaller sets. We do, however, expect that the union of multiple dimensionality reduced sets will not immediately have a high intrinsic dimension. In particular if the data evolves over time, then multiple consecutive  pieces of the input data will have approximately the same best fit subspace (but over time, the subspace will change). We add more flexibility to the algorithm by running more than one copy of BICO, while allowing that more than one SVD output is processed by the same BICO instance. The actual computation tree is visualized in Figure~\ref{fig:computationtreepiecymr}.

\emph{Parameters.} The algorithm has three parameters, the dimension that the SVD reduces to, the \emph{piece size} which is the number of points that are read as input for one SVD computation, and the \emph{number of pieces}, which is the number of SVD outputs that are processed by one instance of BICO. When BICO reaches the limit, the computed coreset is given to a SVD instance and then entered into a BICO on a higher level. It is convenient to set the piece size to $200k$, which also means that BICO computes a summary of size $200k$, the summary size suggested in the original BICO publication.

\emph{Memory requirement.} We store one BICO element for each level of the computation tree. The degree of the tree is equal to the number of pieces $b$, so we have $\log_b n$ levels. At each point in time, there is at most one SVD object in the memory since there is always at most one SVD computation at the same time. If the piece size is equal to $200k$, then the memory requirement of each BICO element is proportional.

\subsubsection{Weighted BICO}

In the original implementation, BICO processes unweighted input points. In the \emph{piecy-mr} computation tree, the instances of BICO on higher levels of the computation tree have to process weighted inputs (since the coreset points are weighted). Thus, we extended the source code of BICO to work for weighted inputs. For an input point $x$ with weight $w$, we have to simulate what BICO would do for $w$ copies of $x$. The main observation is that in most routines of BICO, multiple copies of the same point can be treated as one. For example, finding the closest reference point that is currently in the data structure can be done once and the result is then valid for all copies of $x$. Additionally, if we decide to open a new clustering feature with $x$ as the reference point, we can insert all (not yet inserted) copies into this clustering feature at no cost.

What we have to adjust is the insertion process into already existing clustering features, and the initial values for new clustering features.
Setting the correct values for a new clustering feature is straightforward: The new clustering feature has reference point $x$, its sum of points is $w \cdot x$, the sum of squares is $w \cdot x^2$ and the number of points stored in the feature is $w$. When we add $w$ copies of a point $x$ to an existing clustering feature with centroid $\mu$ and $s$ points in it, then the actual increase of the error due to this is
\begin{align*}
s \cdot \|\mu-\mu_n\|^2 + w \|x-\mu_n\|^2
& = s \cdot \Bigl\|\mu-\frac{s\mu+wx}{s+w}\Bigr\|^2 + w \Bigl\|x-\frac{s\mu+wx}{s+w}\Bigr\|^2 \\
& = \frac{sw^2}{(s+w)^2}\|x-\mu\|^2 + \frac{ws^2}{(s+w)^2}\|x-\mu\|^2\\
& = \frac{sw}{s+w }\|x-\mu\|^2
\end{align*}
where we denote the new centroid after adding $w$ copies of $x$ by $\mu_n$.
We conclude that the total error made in the feature after inserting $w$ points is
$c + \frac{sw}{s+w} \| x- \mu \|^2$, where $c$ denotes the original error made in the feature.

The original BICO implementation would have inserted the $w$ copies sequentially into the clustering feature
until the features threshold error of $T$ would have been surpassed. It actually uses $\|x-\mu\|^2$ to measure the additional error and thus overestimates it. When adding single points, the effect of this overestimation decreases with each added point such that this works well for BICO. In the weighted version, however, using $w \cdot \|x-\mu\|^2$ is can be off by a large margin.

Instead,
we compute how many copies $w'$ of $x$ can be inserted into the
feature without surpassing the threshold:
\begin{align*}
c + \frac{sw'}{s+w'} \| x- \mu \|^2 \leq T \iff\ w' \bigl( s \|x-\mu\|^2 - T + c\bigr) \leq sT - sc
\end{align*}
If $s \cdot \|x-\mu\|^2 - T + c \leq 0$, the threshold will not be reached for any $w' \geq 0$.
We can thus insert all $w$ copies. Otherwise, we insert
\begin{align*}
w' = \min\Big(w, \frac{sT - sc}{s\|x - \mu\|^2 - T + c}\thinspace\Big)
\end{align*}
many copies of $x$. If the threshold is reached before all $w$ copies of $x$ are inserted, \ie if $w' < w$,
we continue recursively as in the original BICO implementation.

\subsubsection{Best fit subspace for weighted points}

The singular value decomposition of a matrix is defined in an unweighted fashion, yet we want to use it for reducing the dimensionality of the weighted coreset points that result from BICO runs. What we want to do is project the points to the best fit subspace of the point set where each point is replaced by several copies of itself according to its weight. Translated into the matrix notation, this means that we want to compute the projection of $A$ to the best fit subspace of dimension $\ell$ of a matrix $F$ which contains multiple copies of the points from $A$ according to their (integral) weight\footnote{The weights that are computed by BICO are always integral. In fact, they sum up to the number of points BICO has processed.}.

Certainly, we do not want to actually create $F$. Instead, we construct a matrix $A'$ where each row $A_{i\ast}$ is replaced by $\sqrt{w_i} A_{i\ast}$ where $w_i$ is the weight of the $i$th point. By linear algebra, we can verify that for each pair of left and right singular vectors $u$ and $v$ of $F$ with singular value $\sigma$, there exists a vector $u'$ such that $u'$ and $v$ are a pair of left and right singular vectors of $A'$ for the same singular value. The reverse direction also holds. Thus, $A'$ and $F$ have the same best fit subspace and we can compute the SVD of $A'$ in order to obtain it. After obtaining $A'_\ell$, we divide each row $i$ by $\sqrt{w_i}$ to get the projection of the points in $A$. Their weight does not change.

Notice that we cannot replace weighted points by some multiplied version when we input the points into BICO since the clustering behaviour of a weighted point differs from the clustering behaviour of any multiple (imagine a center that lies at the weighted point, so that it has no cost -- but any multiplied point would have).

\section{Experiments}
The experiments were performed in three settings. For class I, all source codes were compiled using gcc~4.9.1, and experiments were performed on 20 identical machines with a 3.2~GHz AMD Phenom II\textsuperscript{™} X6 1090T processor and 8~GiB RAM.
For class II, all source codes were compiled with gcc~4.8.2 and all experiments were performed on 7 identical machines with a 2.8~GHz Intel\textsuperscript{®} E7400 processor and 8~GiB RAM.
In class III, all source codes were compiled with gcc~4.9.1 and all experiments were performed on one machine with
a 2.6~GHz Intel\textsuperscript{®} Core™ i5-4210M CPU processor and 16~GiB RAM.

\begin{figure}
\includegraphics[]{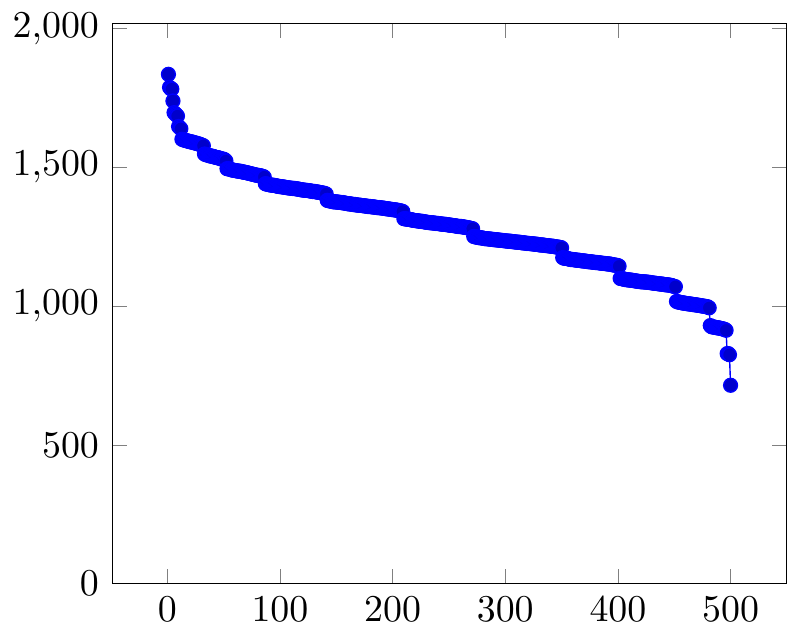}
\includegraphics[]{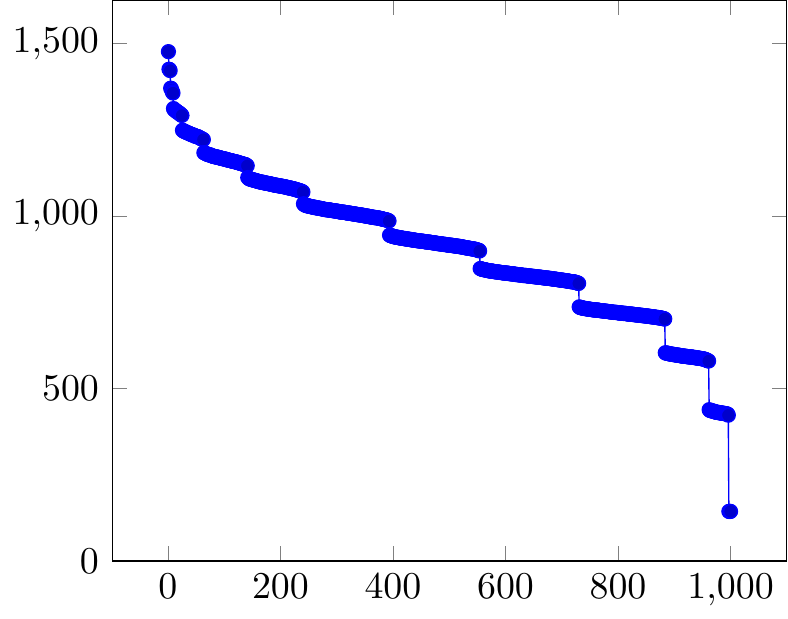}
\caption{Spectrum of two \emph{StructuredWithNoise} data sets with $d=500$ and $d=1000$, containing $50$ clusters of $5000$ points each.\label{spectrumswn}}
\end{figure}

\begin{figure}
\includegraphics[]{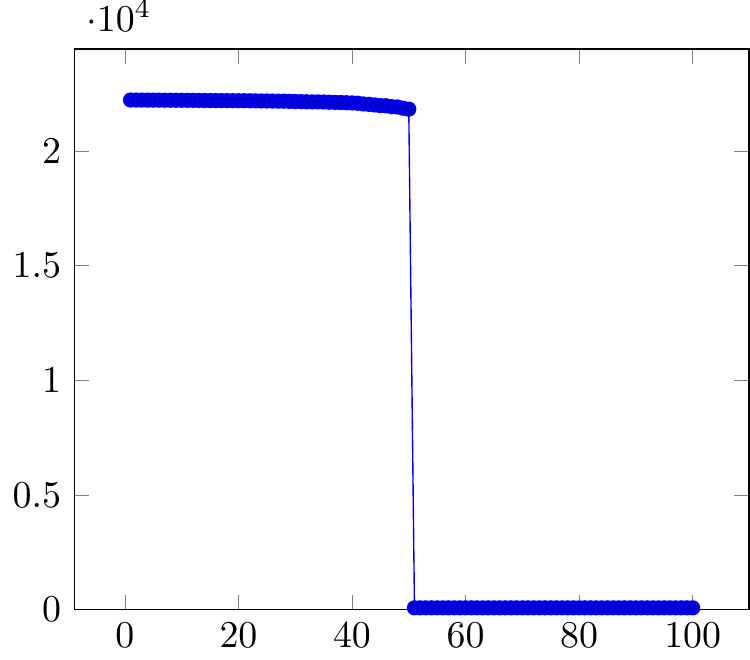}
\includegraphics[]{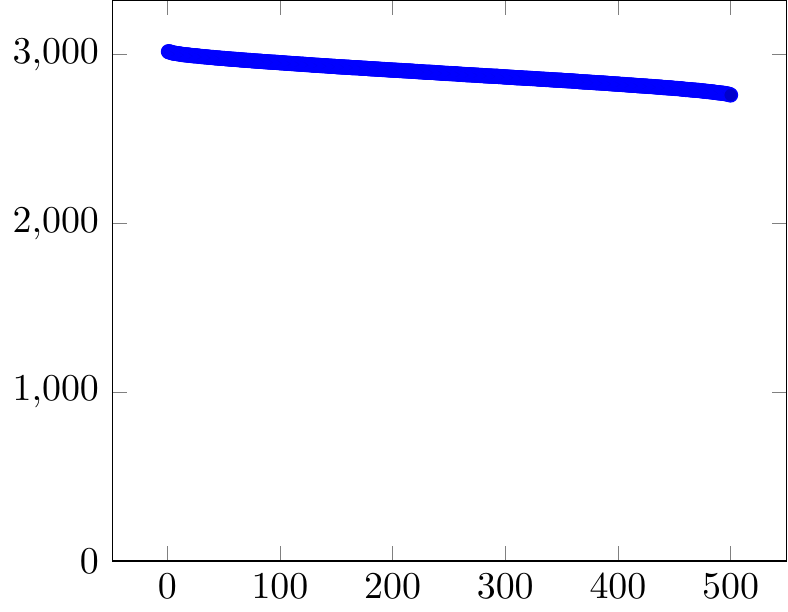}
\caption{Spectrum of a LowerBound data set with $d=10000$ and a random data set with $d=10000$.\label{spectrumlowerbound}}
\end{figure}

Our testbed consists of the following instances. Notice that we computed the spectrum for examples of the data set families. This gives an additional insight on the structure of the data sets.
\begin{description}
\item[Caltech128] The \texttt{Caltech128} instance was created from the Caltech101
image database~\cite{FFFP2004} and consists of 128 SIFT descriptors~\cite{Lowe04},
resulting in 128 dimensions and about 3.1 million points.
The instance was used in~\cite{FGSSS13} for BICO benchmarks and was provided to the
authors by Grzeszick in a private communication.
\item[StructuredWithNoise]
The idea of the \texttt{StructuredWithNoise} instances is to  hide $\ell \in \N$
random point sets of $y \in \N$ points in $\R^{d}$.
To build cluster $i \in \{1,\dots, \ell\}$, select $x$ dimensions
$D_i = \{d_1,\dots,d_x\} \subseteq \{1,\dots,d\}$ uniformly at random.
Then we build the $y$ points for cluster $i$: For point $j$, choose
the coordinates corresponding to $D_i$ uniformly at random from $[-\Delta, \Delta]$.
Select the remaining coordinates, \ie, the noise, uniformly at random
from $[-\delta, \delta]$.
This yields an instance with $\ell\cdot y$ points of dimension $d$.
We fix $\Delta=10$, $\delta=1/2$.
Figure~\ref{spectrumswn} shows the spectrum of two \texttt{StructuredWithNoise} data sets. We see that the first singular values are large, followed by slowly decreasing values until the descent steepens again.

\item[LowerBound]
Arthur and Vassilvitskii~\cite{AV07} propose the following class of worst-case
instances for the \texttt{kmeans++} algorithm.
Define the (affine) $(k,\Delta)$-simplex as the convex combination of the $k$
unit vectors $e_1,\dots,e_k$ in $\R^k$, scaled by $\Delta > 0$.
Now, embed such a $(k,\Delta)$-simplex $\mathfrak{S}$
in the first $k$ dimensions of $\R^{k+n}$.
Then use the remaining $n$ dimensions of $\R^{k+n}$ to place a $(n/k, \delta)$-simplex $S_i$
in each vertex $i$ of $\mathfrak{S}$ such that all $S_i$ use disjoint dimensions.
Arthur and Vassilvitskii~\cite{AV07} prove that the \texttt{kmeans++} algorithm
can achieve no better approximation ratio then $\Omega(\log N)$ on this class of instances,
where $N$ is the number of input points. We use a generator by Stallmann~\cite{Stallmann2014}
to generate instance of this type. We fix $\delta=100$ and $\Delta=1000$.
The \texttt{LowerBound} data sets have a nice structure for our experiments since the only the first singular values are significant as can be seen in the left diagram in Figure~\ref{spectrumlowerbound}. Notice that we computed the first 100 singular values for a $10000$-dimensional data set. The remaining values can only be smaller.
\item[Random]
A \texttt{Random} data set is created by computing $n^2$ random numbers from $[-\Delta,\Delta]$ to form an $n$-dimensional data set with $n$ points. We used $\Delta = 10$. Notice that the expected directional width is not equal for all directions (the points are drawn uniformly from a cube, not from a sphere). The resulting spectrum is slightly decreasing (see Figure~\ref{spectrumlowerbound}, right diagram).

\end{description}

Since the algorithms are randomized, we repeated all experiments five times with the exception of the the test cases for the three largest \texttt{StructuredWithNoise} data sets because of computation times.

\subsection{redSVD as a replacement for the lapack++ SVD}
Replacing the exact SVD computation in our algorithm by an approximative one as outlined in Section~\ref{sec:the-algorithms} can only work if the approximation is fast and provides reliable results.

Additionally, we are interested in the factor of speed that can be gained by switching to redSVD from a full SVD computation.

To evaluate the redSVD performance, we use a test bed of \texttt{StructuredWithNoise} instances with varying values for $y$ and $d$ thus yielding instances from small to huge size. The results are depicted in Table~\ref{table:redsvd}.

We use redSVD to replace the input $A$ by a matrix $A_\ell'$. To measure the error of redSVD, we compare $||A-A_\ell'||_F^2$ to $||A-A_\ell||_F^2$, where $A_\ell$ is the matrix computed by the full SVD implementation in lapack++. The matrix obtained by projecting $A$ to its best fit subspace of dimension $\ell$ minimizes the Frobenius norm of the difference to $A$, so this is a suitable measure to evaluate the redSVD result. The table shows the deviation of redSVD compared to the Frobenius distance of the matrix computed by the full SVD.

We performed the SVD comparison reducing the dimension to values in $\{100, 125, 150\}$.

We found that the error made by redSVD is indeed very small (less than 7\% in all cases) while
computation times become significantly faster: instances with 30,000 rows in 1000 columns
can still be solved by redsvd in about 3s while \texttt{lapack++}'s takes 3000s on the same instace.
RedSVD was able to compute approximate SVDs of matrices with 500,000 rows and 500 columns in 40s.

The limiting factor to solve larger instances is in both cases the memory limitation.
The largest instance that we could compute full SVD on was with contains $n=30000$ points in $d=1000$ dimensions (constructed with $y=300$ and $k=100$). The redSVD approach uses much smaller matrices and thus it is possible to solve \texttt{StructureWithNoise} instances up to $n=500000$ and $d=500$. Then, however, it also stops working.
Observe that computing the redSVD on this $500.000 \times 500$ matrix is still faster than one computation of a full SVD for instances with size $n=10000$ and $d=500$.

\sisetup{
  scientific-notation=false,
	round-mode = places,
	round-precision=2,
}

{\scriptsize%
\begin{longtable}{l@{\hskip 12\tabcolsep}S[table-format=1.2]@{\hspace{8\tabcolsep}}S[table-format=1.2]@{\hskip 8\tabcolsep}S[table-format=1.2]@{\hskip 8\tabcolsep}S[table-format=1.2]@{\hskip 12\tabcolsep}S[table-format=4]@{\hskip 6\tabcolsep}S[table-format=4]@{\hskip 6\tabcolsep}S[table-format=4]@{\hskip 12\tabcolsep}S[table-format=2]@{\hskip 6\tabcolsep}S[table-format=2]@{\hskip 6\tabcolsep}S[table-format=2]}
\toprule
Group & \multicolumn{4}{@{}c@{\hskip 12\tabcolsep}}{Percent error} &  \multicolumn{3}{@{}c@{\hskip 12\tabcolsep}}{Full SVD CPU }& \multicolumn{3}{c}{redSVD CPU} \\
\cmidrule(lr{22.4pt}){2-5} %
\cmidrule(lr{22.4pt}){6-8}
\cmidrule(lr){9-11}
{SWN, $k=100$} &
{min} &  {max} &  {avg} &  {med} &
{min} &  {max} &  {avg} &
{min} &  {max} &  {avg} \\
\midrule \endfirsthead
{SWN, $k=100$} &
{min} &  {max} &  {avg} &  {med} &
{min} &  {max} &  {avg} &
{min} &  {max} &  {avg} \\
\midrule \endhead
\bottomrule
\caption{Comparison of the full SVD by lapack++ with redSVD.}\\%
\endfoot
\bottomrule
\caption[]{Comparison of the full SVD by lapack++ with redSVD on various randomized instances with $k=100$ and varying parameters. The table shows error percentage of the approximate solution and the running times in seconds. Notice that the number of points in the instances is $k\cdot y$. Experiment belongs to class I.
Given a matrix $A$, its full SVD $A_\ell$ and its approximate SVD $A_\ell'$, we verify the accuracy
of the redSVD approximation by comparing $||A-A_\ell'||_F^2$ to $||A-A_\ell||_F^2$ on instances
of the \texttt{StructureWithNoise} class.
The matrix obtained by projecting $A$ to its best fit subspace of dimension $\ell$ minimizes the Frobenius norm of the difference to $A$, so this is a suitable measure to evaluate the redSVD result.
\label{table:redsvd}}\\%
\endlastfoot
y-100-d-500-svd-100                        &
0.5203 & 0.7915 & 0.6639 & 0.6837 & 167 &  169 &  168 &  0 &  0 &  0 \\
y-100-d-500-svd-125                        &
-2.9118 & -2.6788 & -2.7603 & -2.7457 & 167 &  169 &  168 &  0 &  0 &  0 \\
y-100-d-500-svd-150                        &
-6.4092 & -6.1900 & -6.2741 & -6.2547 & 167 &  169 &  168 &  0 &  0 &  0 \\
y-100-d-1000-svd-100                       &
1.3805 & 1.5140 & 1.4440 & 1.4532 & 350 &  353 &  351 &  0 &  0 &  0 \\
y-100-d-1000-svd-125                       &
-0.2875 & -0.1312 & -0.2106 & -0.2067 & 350 &  353 &  351 &  0 &  0 &  0 \\
y-100-d-1000-svd-150                       &
-1.9612 & -1.8157 & -1.8864 & -1.8780 & 350 &  353 &  351 &  0 &  1 &  1 \\
y-200-d-500-svd-100                        &
0.0130 & 0.1634 & 0.0977 & 0.1133 & 657 &  663 &  659 &  0 &  0 &  0 \\
y-200-d-500-svd-125                        &
-3.3530 & -3.1387 & -3.2448 & -3.2496 & 657 &  663 &  659 &  1 &  1 &  1 \\
y-200-d-500-svd-150                        &
-6.7721 & -6.5690 & -6.6623 & -6.6652 & 657 &  663 &  659 &  1 &  1 &  1 \\
y-200-d-1000-svd-100                       &
0.9987 & 1.1605 & 1.0624 & 1.0436 & 1347 & 1356 & 1351 &  1 &  1 &  1 \\
y-200-d-1000-svd-125                       &
-0.5825 & -0.4313 & -0.5144 & -0.5148 & 1347 & 1356 & 1351 &  1 &  1 &  1 \\
y-200-d-1000-svd-150                       &
-2.1661 & -2.0625 & -2.1190 & -2.1142 & 1347 & 1356 & 1351 &  1 &  2 &  2 \\
y-300-d-500-svd-100                        &
-0.1695 & -0.0520 & -0.1266 & -0.0838 & 1480 & 1485 & 1483 &  1 &  1 &  1 \\
y-300-d-500-svd-125                        &
-3.5114 & -3.4067 & -3.4639 & -3.4350 & 1480 & 1485 & 1483 &  1 &  1 &  1 \\
y-300-d-500-svd-150                        &
-6.8759 & -6.7812 & -6.8397 & -6.8138 & 1480 & 1485 & 1483 &  1 &  2 &  2 \\
y-300-d-1000-svd-100                       &
0.9119 & 0.9347 & 0.9231 & 0.9299 & 3028 & 3039 & 3034 &  2 &  2 &  2 \\
y-300-d-1000-svd-125                       &
-0.6637 & -0.6273 & -0.6459 & -0.6323 & 3028 & 3039 & 3034 &  2 &  2 &  2 \\
y-300-d-1000-svd-150                       &
-2.2472 & -2.2115 & -2.2216 & -2.2126 & 3028 & 3039 & 3034 &  3 &  3 &  3 \\
y-500-d-500-svd-100                        &
\textemdash & \textemdash & \textemdash & \textemdash & \textemdash & \textemdash & \textemdash &   2 & 2 &  2 \\
y-500-d-500-svd-125                        &
\textemdash & \textemdash & \textemdash & \textemdash & \textemdash & \textemdash & \textemdash &   2 &  2 &  2 \\
y-500-d-500-svd-150                        &
\textemdash & \textemdash & \textemdash & \textemdash & \textemdash & \textemdash & \textemdash &   3 &  3 &  3 \\
y-500-d-1000-svd-100                       &
\textemdash & \textemdash & \textemdash & \textemdash & \textemdash & \textemdash & \textemdash &   3 &  3 &  3 \\
y-500-d-1000-svd-125                       &
\textemdash & \textemdash & \textemdash & \textemdash & \textemdash & \textemdash & \textemdash &   4 &  4 &  4 \\
y-500-d-1000-svd-150                       &
\textemdash & \textemdash & \textemdash & \textemdash & \textemdash & \textemdash & \textemdash &   4 &  4 &  4 \\
y-1000-d-500-svd-100                       &
\textemdash & \textemdash & \textemdash & \textemdash & \textemdash & \textemdash & \textemdash &   4 &  4 &  4 \\
y-1000-d-500-svd-125                       &
\textemdash & \textemdash & \textemdash & \textemdash & \textemdash & \textemdash & \textemdash &   5 &  5 &  5 \\
y-1000-d-500-svd-150                       &
\textemdash & \textemdash & \textemdash & \textemdash & \textemdash & \textemdash & \textemdash &   6 &  6 &  6 \\
y-1000-d-1000-svd-100                      &
\textemdash & \textemdash & \textemdash & \textemdash & \textemdash & \textemdash & \textemdash &   7 &  7 &  7 \\
y-1000-d-1000-svd-125                      &
\textemdash & \textemdash & \textemdash & \textemdash & \textemdash & \textemdash & \textemdash &   8 &  8 &  8 \\
y-1000-d-1000-svd-150                      &
\textemdash & \textemdash & \textemdash & \textemdash & \textemdash & \textemdash & \textemdash &  10 & 10 & 10 \\
y-2000-d-500-svd-100                       &
\textemdash & \textemdash & \textemdash & \textemdash & \textemdash & \textemdash & \textemdash &   9 &  9 &  9 \\
y-2000-d-500-svd-125                       &
\textemdash & \textemdash & \textemdash & \textemdash & \textemdash & \textemdash & \textemdash &  11 & 11 & 11 \\
y-2000-d-500-svd-150                       &
\textemdash & \textemdash & \textemdash & \textemdash & \textemdash & \textemdash & \textemdash &  13 & 13 & 13 \\
y-2000-d-1000-svd-100                      &
\textemdash & \textemdash & \textemdash & \textemdash & \textemdash & \textemdash & \textemdash &  16 & 16 & 16 \\
y-2000-d-1000-svd-125                      &
\textemdash & \textemdash & \textemdash & \textemdash & \textemdash & \textemdash & \textemdash &  20 & 20 & 20 \\
y-2000-d-1000-svd-150                      &
\textemdash & \textemdash & \textemdash & \textemdash & \textemdash & \textemdash & \textemdash &  23 & 23 & 23 \\
y-5000-d-500-svd-100                       &
\textemdash & \textemdash & \textemdash & \textemdash & \textemdash & \textemdash & \textemdash &  24 & 24 & 24 \\
y-5000-d-500-svd-125                       &
\textemdash & \textemdash & \textemdash & \textemdash & \textemdash & \textemdash & \textemdash &  29 & 29 & 29 \\
y-5000-d-500-svd-150                       &
\textemdash & \textemdash & \textemdash & \textemdash & \textemdash & \textemdash & \textemdash &  36 & 36 & 36 \\
y-5000-d-1000-svd-100                      &
\textemdash & \textemdash & \textemdash & \textemdash & \textemdash & \textemdash & \textemdash & \textemdash & \textemdash & \textemdash \\
y-5000-d-1000-svd-125                      &
\textemdash & \textemdash & \textemdash & \textemdash & \textemdash & \textemdash & \textemdash & \textemdash & \textemdash & \textemdash \\
y-5000-d-1000-svd-150                      &
\textemdash & \textemdash & \textemdash & \textemdash & \textemdash & \textemdash & \textemdash & \textemdash & \textemdash & \textemdash \\
y-10000-d-500-svd-100                      &
\textemdash & \textemdash & \textemdash & \textemdash & \textemdash & \textemdash & \textemdash & \textemdash & \textemdash & \textemdash \\
y-10000-d-500-svd-125                      &
\textemdash & \textemdash & \textemdash & \textemdash & \textemdash & \textemdash & \textemdash & \textemdash & \textemdash & \textemdash \\
y-10000-d-500-svd-150                      &
\textemdash & \textemdash & \textemdash & \textemdash & \textemdash & \textemdash & \textemdash & \textemdash & \textemdash &\textemdash \\
y-10000-d-1000-svd-100                     &
\textemdash & \textemdash & \textemdash & \textemdash & \textemdash & \textemdash & \textemdash & \textemdash & \textemdash & \textemdash \\
\end{longtable}
}

\sisetup{
  scientific-notation=true,
	round-mode = places,
	scientific-notation = fixed,
	round-precision=2,
}

\subsection{Performance of BICO, Piecy and Piecy-MR}

\subsubsection*{BICO.} Table~\ref{bico-experimente} contains the basic test cases and reports the results that BICO achieved when run on the test case directly. Notice that we use the current version of the source code from the BICO website. In contrast to the version used in~\cite{FSS13}, this version has varying running times. This shows both in the BICO experiments itself as in the experiments for \emph{piecy} and \emph{piecy-mr} since they both use BICO. For example, consider the varying running time of BICO on the \texttt{enron} data set. Obviously, \emph{piecy} and \emph{piecy-mr} will improve when the source code of BICO is updated. For this reason, we will pay most attention to the median of the running times and not the average running time.

In all tables, the parameters are listed in the caption if they are equal for all test cases in the table, or at the start of each line if they vary. We denote the number of points by $n$, the dimension by $d$ and the number of centers by $k$.

{ %
\scriptsize%
\begin{longtable}{l@{\hskip 4\tabcolsep}r@{\hskip 4\tabcolsep}r@{\hskip 4\tabcolsep}r@{\hskip 4\tabcolsep}r@{\hskip 8\tabcolsep}r@{\hskip 4\tabcolsep}r@{\hskip 4\tabcolsep}r@{\hskip 4\tabcolsep}r}
\toprule
{Group} & \multicolumn{4}{@{}c@{\hskip 8\tabcolsep}}{Cost} & \multicolumn{4}{c}{Running time}\\
\cmidrule(lr{16.8pt}){2-5} %
\cmidrule(lr){6-9}
{} &
\multicolumn{1}{c@{\hskip 0\tabcolsep}}{min} & \multicolumn{1}{c@{\hskip 4\tabcolsep}}{max} & \multicolumn{1}{c@{\hskip 4\tabcolsep}}{average} & \multicolumn{1}{c@{\hskip 12\tabcolsep}}{median} &
min & max & \multicolumn{1}{c@{\hskip 4\tabcolsep}}{avg} & \multicolumn{1}{c}{med} \\
\midrule \endfirsthead
{} &
\multicolumn{1}{c@{\hskip 4\tabcolsep}}{min} & \multicolumn{1}{c@{\hskip 4\tabcolsep}}{max} & \multicolumn{1}{c@{\hskip 4\tabcolsep}}{average} & \multicolumn{1}{c@{\hskip 12\tabcolsep}}{median} &
min & max & \multicolumn{1}{c@{\hskip 4\tabcolsep}}{avg} & \multicolumn{1}{c}{med} \\
\midrule \endhead
\bottomrule
\caption{BICO (continued)}\\%
\endfoot
\bottomrule
\caption[]{BICO results.\label{bico-experimente}}\\%
\endlastfoot
\multicolumn{3}{l}{\texttt{LowerBound}, experiments belong to class II}\\
\midrule
k-10-n-$10^4$-d-10010%
&   \num[fixed-exponent = 7]{4995e+4} &   \num[fixed-exponent = 7]{4995e+4} & \num[fixed-exponent = 7]{4995e+4} &   \num[fixed-exponent = 7]{4995e+4} &         \num[scientific-notation=false,round-mode=places,round-precision=1]{74} &         \num[scientific-notation=false,round-mode=places,round-precision=1]{77} &       \num[scientific-notation=false,round-mode=places,round-precision=1]{75.6} &         \num[scientific-notation=false,round-mode=places,round-precision=1]{76} \\
k-50-n-$10^4$-d-10050%
&   \num[fixed-exponent = 7]{4975e+4} &  \num[fixed-exponent = 7]{14875e+4} & \num[fixed-exponent = 7]{8935e+4} &   \num[fixed-exponent = 7]{4975e+4} &         \num[scientific-notation=false,round-mode=places,round-precision=1]{78} &         \num[scientific-notation=false,round-mode=places,round-precision=1]{79} &       \num[scientific-notation=false,round-mode=places,round-precision=1]{78.7} &         \num[scientific-notation=false,round-mode=places,round-precision=1]{79} \\
\bottomrule
\multicolumn{3}{l}{\texttt{BagOfWords}, experiments belong to class II}\\
\midrule
enron-k-10                              &   \num[fixed-exponent = 7]{16274100} &   \num[fixed-exponent = 7]{16873600} & \num[fixed-exponent = 7]{16519640.0} &   \num[fixed-exponent = 7]{16564850} &        \num[scientific-notation=false,round-mode=places,round-precision=1]{480} &       \num[scientific-notation=false,round-mode=places,round-precision=1]{1679} &      \num[scientific-notation=false,round-mode=places,round-precision=1]{611.9} &        \num[scientific-notation=false,round-mode=places,round-precision=1]{491} \\
kos-k-2                                 &     \num[fixed-exponent = 5]{389702} &     \num[fixed-exponent = 5]{394916} &   \num[fixed-exponent = 5]{391874.0} &     \num[fixed-exponent = 5]{390699} &         \num[scientific-notation=false,round-mode=places,round-precision=1]{10} &         \num[scientific-notation=false,round-mode=places,round-precision=1]{11} &       \num[scientific-notation=false,round-mode=places,round-precision=1]{10.9} &         \num[scientific-notation=false,round-mode=places,round-precision=1]{11} \\
\bottomrule
\multicolumn{3}{l}{\texttt{Caltech128}, experiments belong to class I}\\
\midrule
k-5                                                            & \num[fixed-exponent = 11]{42336900e+4} & \num[fixed-exponent = 11]{42336900e+4} & \num[fixed-exponent = 11]{42336900e+4} & \num[fixed-exponent = 11]{42336900e+4} & \num[scientific-notation=false,round-mode=places,round-precision=1]{       319} & \num[scientific-notation=false,round-mode=places,round-precision=1]{       319} & \num[scientific-notation=false,round-mode=places,round-precision=1]{     319.1} & \num[scientific-notation=false,round-mode=places,round-precision=1]{       319} \\
k-10                                                           & \num[fixed-exponent = 11]{41299200e+4} & \num[fixed-exponent = 11]{41299200e+4} & \num[fixed-exponent = 11]{41299200e+4} & \num[fixed-exponent = 11]{41299200e+4} & \num[scientific-notation=false,round-mode=places,round-precision=1]{       366} & \num[scientific-notation=false,round-mode=places,round-precision=1]{       366} & \num[scientific-notation=false,round-mode=places,round-precision=1]{     366.0} & \num[scientific-notation=false,round-mode=places,round-precision=1]{       366} \\
k-50                                                           & \num[fixed-exponent = 11]{34279900e+4} & \num[fixed-exponent = 11]{34279900e+4} & \num[fixed-exponent = 11]{34279900e+4} & \num[fixed-exponent = 11]{34279900e+4} & \num[scientific-notation=false,round-mode=places,round-precision=1]{       428} & \num[scientific-notation=false,round-mode=places,round-precision=1]{       428} & \num[scientific-notation=false,round-mode=places,round-precision=1]{     427.6} & \num[scientific-notation=false,round-mode=places,round-precision=1]{       428} \\
k-100                                                          & \num[fixed-exponent = 11]{30428500e+4} & \num[fixed-exponent = 11]{30428500e+4} & \num[fixed-exponent = 11]{30428500e+4} & \num[fixed-exponent = 11]{30428500e+4} & \num[scientific-notation=false,round-mode=places,round-precision=1]{       503} & \num[scientific-notation=false,round-mode=places,round-precision=1]{       503} & \num[scientific-notation=false,round-mode=places,round-precision=1]{     502.9} & \num[scientific-notation=false,round-mode=places,round-precision=1]{       503} \\
k-250                                                          & \num[fixed-exponent = 11]{27389400e+4} & \num[fixed-exponent = 11]{27389400e+4} & \num[fixed-exponent = 11]{27389400e+4} & \num[fixed-exponent = 11]{27389400e+4} & \num[scientific-notation=false,round-mode=places,round-precision=1]{       571} & \num[scientific-notation=false,round-mode=places,round-precision=1]{       571} & \num[scientific-notation=false,round-mode=places,round-precision=1]{     571.1} & \num[scientific-notation=false,round-mode=places,round-precision=1]{       571} \\
k-1000                                                         & \num[fixed-exponent = 11]{23437100e+4} & \num[fixed-exponent = 11]{23437100e+4} & \num[fixed-exponent = 11]{23437100e+4} & \num[fixed-exponent = 11]{23437100e+4} & \num[scientific-notation=false,round-mode=places,round-precision=1]{       560} & \num[scientific-notation=false,round-mode=places,round-precision=1]{       560} & \num[scientific-notation=false,round-mode=places,round-precision=1]{     559.7} & \num[scientific-notation=false,round-mode=places,round-precision=1]{       560} \\
\bottomrule
\multicolumn{3}{l}{\texttt{Random}, experiments belong to class II}\\
\midrule
n-$10^6$-d-1000-k-10               & \num[fixed-exponent = 10]{3328190e+4} & \num[fixed-exponent = 10]{3329500e+4} & \num[fixed-exponent = 10]{3328656e+4} & \num[fixed-exponent = 10]{3328410e+4} & \num[scientific-notation=false,round-mode=places,round-precision=1]{      1058} & \num[scientific-notation=false,round-mode=places,round-precision=1]{      2126} & \num[scientific-notation=false,round-mode=places,round-precision=1]{    1718.6} & \num[scientific-notation=false,round-mode=places,round-precision=1]{      1816} \\
n-$10^6$-d-1000-k-20               & \num[fixed-exponent = 10]{3327390e+4} & \num[fixed-exponent = 10]{3328560e+4} & \num[fixed-exponent = 10]{3328188e+4} & \num[fixed-exponent = 10]{3328310e+4} & \num[scientific-notation=false,round-mode=places,round-precision=1]{      2578} & \num[scientific-notation=false,round-mode=places,round-precision=1]{      4792} & \num[scientific-notation=false,round-mode=places,round-precision=1]{    3522.8} & \num[scientific-notation=false,round-mode=places,round-precision=1]{      2952} \\
n-$10^6$-d-1000-k-50               & \num[fixed-exponent = 10]{3325910e+4} & \num[fixed-exponent = 10]{3328630e+4} & \num[fixed-exponent = 10]{3327082e+4} & \num[fixed-exponent = 10]{3326970e+4} & \num[scientific-notation=false,round-mode=places,round-precision=1]{      1004} & \num[scientific-notation=false,round-mode=places,round-precision=1]{      4466} & \num[scientific-notation=false,round-mode=places,round-precision=1]{    2326.8} & \num[scientific-notation=false,round-mode=places,round-precision=1]{      1819} \\
\bottomrule
\multicolumn{3}{l}{\texttt{StructuredWithNoise}, experiments belong to class I}\\
\midrule
y-5000-d-1000-k-10                & \num[fixed-exponent = 9]{170239e+4} & \num[fixed-exponent = 9]{170408e+4} & \num[fixed-exponent = 9]{170341e+4} & \num[fixed-exponent = 9]{170346e+4} & \num[scientific-notation=false,round-mode=places,round-precision=1]{       368} & \num[scientific-notation=false,round-mode=places,round-precision=1]{      1227} & \num[scientific-notation=false,round-mode=places,round-precision=1]{     610.1} & \num[scientific-notation=false,round-mode=places,round-precision=1]{       592} \\
y-5000-d-1000-k-20                & \num[fixed-exponent = 9]{170219e+4} & \num[fixed-exponent = 9]{170408e+4} & \num[fixed-exponent = 9]{170337e+4} & \num[fixed-exponent = 9]{170329e+4} & \num[scientific-notation=false,round-mode=places,round-precision=1]{       591} & \num[scientific-notation=false,round-mode=places,round-precision=1]{      2204} & \num[scientific-notation=false,round-mode=places,round-precision=1]{    1217.8} & \num[scientific-notation=false,round-mode=places,round-precision=1]{      1085} \\
y-5000-d-1000-k-50                & \num[fixed-exponent = 9]{169772e+4} & \num[fixed-exponent = 9]{170233e+4} & \num[fixed-exponent = 9]{169956e+4} & \num[fixed-exponent = 9]{169950e+4} & \num[scientific-notation=false,round-mode=places,round-precision=1]{       547} & \num[scientific-notation=false,round-mode=places,round-precision=1]{      2865} & \num[scientific-notation=false,round-mode=places,round-precision=1]{    1133.8} & \num[scientific-notation=false,round-mode=places,round-precision=1]{       872} \\
y-5000-d-1000-k-100               & \num[fixed-exponent = 9]{169279e+4} & \num[fixed-exponent = 9]{170103e+4} & \num[fixed-exponent = 9]{169674e+4} & \num[fixed-exponent = 9]{169618e+4} & \num[scientific-notation=false,round-mode=places,round-precision=1]{       714} & \num[scientific-notation=false,round-mode=places,round-precision=1]{      7679} & \num[scientific-notation=false,round-mode=places,round-precision=1]{    2275.1} & \num[scientific-notation=false,round-mode=places,round-precision=1]{      1359} \\
y-10000-d-500-k-10                & \num[fixed-exponent = 9]{335634e+4} & \num[fixed-exponent = 9]{336594e+4} & \num[fixed-exponent = 9]{336296e+4} & \num[fixed-exponent = 9]{336400e+4} & \num[scientific-notation=false,round-mode=places,round-precision=1]{       411} & \num[scientific-notation=false,round-mode=places,round-precision=1]{      1284} & \num[scientific-notation=false,round-mode=places,round-precision=1]{     740.9} & \num[scientific-notation=false,round-mode=places,round-precision=1]{       691} \\
y-10000-d-500-k-20                & \num[fixed-exponent = 9]{335170e+4} & \num[fixed-exponent = 9]{336594e+4} & \num[fixed-exponent = 9]{336221e+4} & \num[fixed-exponent = 9]{336498e+4} & \num[scientific-notation=false,round-mode=places,round-precision=1]{       435} & \num[scientific-notation=false,round-mode=places,round-precision=1]{      2392} & \num[scientific-notation=false,round-mode=places,round-precision=1]{    1030.0} & \num[scientific-notation=false,round-mode=places,round-precision=1]{       805} \\
y-10000-d-500-k-50                & \num[fixed-exponent = 9]{333576e+4} & \num[fixed-exponent = 9]{336594e+4} & \num[fixed-exponent = 9]{335560e+4} & \num[fixed-exponent = 9]{335925e+4} & \num[scientific-notation=false,round-mode=places,round-precision=1]{       576} & \num[scientific-notation=false,round-mode=places,round-precision=1]{      4772} & \num[scientific-notation=false,round-mode=places,round-precision=1]{    2295.2} & \num[scientific-notation=false,round-mode=places,round-precision=1]{      2084} \\
y-10000-d-500-k-100               & \num[fixed-exponent = 9]{332466e+4} & \num[fixed-exponent = 9]{336594e+4} & \num[fixed-exponent = 9]{334047e+4} & \num[fixed-exponent = 9]{333754e+4} & \num[scientific-notation=false,round-mode=places,round-precision=1]{       846} & \num[scientific-notation=false,round-mode=places,round-precision=1]{      6233} & \num[scientific-notation=false,round-mode=places,round-precision=1]{    2168.6} & \num[scientific-notation=false,round-mode=places,round-precision=1]{      1434} \\
y-10000-d-1000-k-10               & \num[fixed-exponent = 9]{340603e+4} & \num[fixed-exponent = 9]{340786e+4} & \num[fixed-exponent = 9]{340702e+4} & \num[fixed-exponent = 9]{340674e+4} & \num[scientific-notation=false,round-mode=places,round-precision=1]{       722} & \num[scientific-notation=false,round-mode=places,round-precision=1]{      2669} & \num[scientific-notation=false,round-mode=places,round-precision=1]{    1454.9} & \num[scientific-notation=false,round-mode=places,round-precision=1]{      1244} \\
y-10000-d-1000-k-20               & \num[fixed-exponent = 9]{340383e+4} & \num[fixed-exponent = 9]{340786e+4} & \num[fixed-exponent = 9]{340560e+4} & \num[fixed-exponent = 9]{340536e+4} & \num[scientific-notation=false,round-mode=places,round-precision=1]{       770} & \num[scientific-notation=false,round-mode=places,round-precision=1]{      4521} & \num[scientific-notation=false,round-mode=places,round-precision=1]{    2350.0} & \num[scientific-notation=false,round-mode=places,round-precision=1]{      2230} \\
y-10000-d-1000-k-50               & \num[fixed-exponent = 9]{339942e+4} & \num[fixed-exponent = 9]{340786e+4} & \num[fixed-exponent = 9]{340425e+4} & \num[fixed-exponent = 9]{340468e+4} & \num[scientific-notation=false,round-mode=places,round-precision=1]{      1299} & \num[scientific-notation=false,round-mode=places,round-precision=1]{      8648} & \num[scientific-notation=false,round-mode=places,round-precision=1]{    4547.8} & \num[scientific-notation=false,round-mode=places,round-precision=1]{      4897} \\
y-10000-d-1000-k-100              & \num[fixed-exponent = 9]{338719e+4} & \num[fixed-exponent = 9]{340786e+4} & \num[fixed-exponent = 9]{339515e+4} & \num[fixed-exponent = 9]{339508e+4} & \num[scientific-notation=false,round-mode=places,round-precision=1]{      1477} & \num[scientific-notation=false,round-mode=places,round-precision=1]{      8626} & \num[scientific-notation=false,round-mode=places,round-precision=1]{    3602.6} & \num[scientific-notation=false,round-mode=places,round-precision=1]{      2605} \\%
\bottomrule
\multicolumn{3}{l}{\texttt{StructuredWithNoise}, experiments belong to class III}\\
\midrule
y-1000000-d-500-k-50              & \num[fixed-exponent=9]{333837e+4} & \num[fixed-exponent=9]{335024e+4} & \num[fixed-exponent=9]{334439e+4} & \num[fixed-exponent=9]{334495e+4} & \num[scientific-notation=false,round-mode=places,round-precision=1]{       335} & \num[scientific-notation=false,round-mode=places,round-precision=1]{      4192} & \num[scientific-notation=false,round-mode=places,round-precision=1]{    1280.2} & \num[scientific-notation=false,round-mode=places,round-precision=1]{       938} \\
\end{longtable}
}

\subsubsection*{Piecy.}
For piecy, we test the influence of two parameters, the piece size, abbreviation \emph{ps}, and the number of dimensions to which we project the points, abbreviation \emph{svd}. We computed an extensive number of test cases for the data set \texttt{CalTech128} to study the influence of the parameters. Table~\ref{piecy-caltech} summarizes the results for piecy. For $k=5,10,50$, piecy is \emph{always} faster than BICO. The table shows that larger values of $svd$ increase the running time, which is expected, but stays below the running time of BICO for these test cases. The accuracy of piecy is high, in particular for larger svd values. At $k=100$, the situation starts to change as there are three test cases where piecy is slower than BICO. For $k=250,1000$ the results by piecy become somewhat unpredictable. Notice that the number of centers is here higher than the input dimension of the points (which is 128). Thus, piecy cannot gain anything from projecting to a number of dimensions $\geq k$, and the SVD processing becomes overhead. It is thus clear that piecy does not perform as well on these test cases.

{%
\scriptsize%
\begin{longtable}{l@{\hskip 4\tabcolsep}r@{\hskip 4\tabcolsep}r@{\hskip 4\tabcolsep}r@{\hskip 4\tabcolsep}r@{\hskip 8\tabcolsep}r@{\hskip 4\tabcolsep}r@{\hskip 4\tabcolsep}r@{\hskip 4\tabcolsep}r}
\toprule
{Group} & \multicolumn{4}{@{}c@{\hskip 8\tabcolsep}}{Cost} & \multicolumn{4}{c}{Running time}\\
\cmidrule(lr{16.8pt}){2-5} %
\cmidrule(lr){6-9}
{} &
\multicolumn{1}{c@{\hskip 4\tabcolsep}}{min} & \multicolumn{1}{c@{\hskip 4\tabcolsep}}{max} & \multicolumn{1}{c@{\hskip 4\tabcolsep}}{average} & \multicolumn{1}{c@{\hskip 12\tabcolsep}}{median} &
min & max & \multicolumn{1}{c@{\hskip 4\tabcolsep}}{avg} & \multicolumn{1}{c}{med} \\
\midrule \endfirsthead
{} &
\multicolumn{1}{c@{\hskip 4\tabcolsep}}{min} & \multicolumn{1}{c@{\hskip 4\tabcolsep}}{max} & \multicolumn{1}{c@{\hskip 4\tabcolsep}}{average} & \multicolumn{1}{c@{\hskip 12\tabcolsep}}{median} &
min & max & \multicolumn{1}{c@{\hskip 4\tabcolsep}}{avg} & \multicolumn{1}{c}{med} \\
\midrule \endhead
\bottomrule
\caption{Piecy on \texttt{Caltech128}, experiments belong to class I.}\\%
\endfoot
\bottomrule
\caption[]{Piecy on \texttt{Caltech128} (continued), experiments belong to class I.\label{piecy-caltech}}\\%
\endlastfoot
{$k=5$}\\
\midrule
ps-1000-s-10                                         & \num[fixed-exponent = 11]{43684100e+4} & \num[fixed-exponent = 11]{46475600e+4} & \num[fixed-exponent = 11]{45162180e+4} & \num[fixed-exponent = 11]{45359300e+4} & \num[scientific-notation=false,round-mode=places,round-precision=1]{       202} & \num[scientific-notation=false,round-mode=places,round-precision=1]{       244} & \num[scientific-notation=false,round-mode=places,round-precision=1]{     217.7} & \num[scientific-notation=false,round-mode=places,round-precision=1]{       209} \\
ps-1000-s-20                                         & \num[fixed-exponent = 11]{43467300e+4} & \num[fixed-exponent = 11]{44582400e+4} & \num[fixed-exponent = 11]{43819640e+4} & \num[fixed-exponent = 11]{43648100e+4} & \num[scientific-notation=false,round-mode=places,round-precision=1]{       201} & \num[scientific-notation=false,round-mode=places,round-precision=1]{       259} & \num[scientific-notation=false,round-mode=places,round-precision=1]{     228.8} & \num[scientific-notation=false,round-mode=places,round-precision=1]{       235} \\
ps-1000-s-50                                         & \num[fixed-exponent = 11]{42490600e+4} & \num[fixed-exponent = 11]{45149000e+4} & \num[fixed-exponent = 11]{43643540e+4} & \num[fixed-exponent = 11]{43587500e+4} & \num[scientific-notation=false,round-mode=places,round-precision=1]{       224} & \num[scientific-notation=false,round-mode=places,round-precision=1]{       256} & \num[scientific-notation=false,round-mode=places,round-precision=1]{     236.5} & \num[scientific-notation=false,round-mode=places,round-precision=1]{       234} \\
\end{longtable}
}

On the \texttt{Random} instance, piecy performs rather badly. The instance is large (one million points with 1000 dimensions, \ie a total of $10^9$ input numbers). In this case, most of the advantage due to the dimensionality reduction is lost because too many pieces are processed and contribute to the intrinsic dimension of the point set that is given to BICO. A similar behavior can be observed for the three largest \texttt{StructuredWithNoise} data sets. In particular when $n$ reaches a million points, piecys running time goes up.

On the smaller \texttt{LowerBound} test cases though, piecy again outperforms BICO's running time. The \texttt{LowerBound} instances have a huge dimension of $10^5$ but the number of points is also bounded by $10^5$. Thus, there is less time for piecy to accumulate to many intrinsic dimensions.
{%
\scriptsize%
\begin{longtable}{l@{\hskip 4\tabcolsep}r@{\hskip 4\tabcolsep}r@{\hskip 4\tabcolsep}r@{\hskip 4\tabcolsep}r@{\hskip 8\tabcolsep}r@{\hskip 4\tabcolsep}r@{\hskip 4\tabcolsep}r@{\hskip 4\tabcolsep}r}
\toprule
{Group} & \multicolumn{4}{@{}c@{\hskip 8\tabcolsep}}{Cost} & \multicolumn{4}{c}{Running time}\\
\cmidrule(lr{16.8pt}){2-5} %
\cmidrule(lr){6-9}
{} &
\multicolumn{1}{c@{\hskip 4\tabcolsep}}{min} & \multicolumn{1}{c@{\hskip 4\tabcolsep}}{max} & \multicolumn{1}{c@{\hskip 4\tabcolsep}}{average} & \multicolumn{1}{c@{\hskip 12\tabcolsep}}{median} &
min & max & \multicolumn{1}{c@{\hskip 4\tabcolsep}}{avg} & \multicolumn{1}{c}{med} \\
\midrule \endfirsthead
{} &
\multicolumn{1}{c@{\hskip 4\tabcolsep}}{min} & \multicolumn{1}{c@{\hskip 4\tabcolsep}}{max} & \multicolumn{1}{c@{\hskip 4\tabcolsep}}{average} & \multicolumn{1}{c@{\hskip 12\tabcolsep}}{median} &
min & max & \multicolumn{1}{c@{\hskip 4\tabcolsep}}{avg} & \multicolumn{1}{c}{med} \\
\midrule \endhead
\bottomrule
\caption{Piecy on \texttt{Random} instances with $n=10^6$ and $d=1000$ and varying parameters, experiments belong to class II.}\\%
\endfoot
\bottomrule
\caption[]{Piecy on \texttt{Random} instances with $n=10^6$ and $d=1000$ and varying parameters, experiments belong to class II.\label{piecy-random}}\\%
\endlastfoot
{$k=10$}\\
\midrule
ps-2000-svd-10             & \num[fixed-exponent=10]{3326430e+4} & \num[fixed-exponent=10]{3327080e+4} & \num[fixed-exponent=10]{3326748e+4} & \num[fixed-exponent=10]{3326720e+4} & \num[scientific-notation=false,round-mode=places,round-precision=1]{      1013} & \num[scientific-notation=false,round-mode=places,round-precision=1]{      1464} & \num[scientific-notation=false,round-mode=places,round-precision=1]{    1283.2} & \num[scientific-notation=false,round-mode=places,round-precision=1]{      1289} \\
ps-2000-svd-20             & \num[fixed-exponent=10]{3328270e+4} & \num[fixed-exponent=10]{3330310e+4} & \num[fixed-exponent=10]{3329178e+4} & \num[fixed-exponent=10]{3328950e+4} & \num[scientific-notation=false,round-mode=places,round-precision=1]{       997} & \num[scientific-notation=false,round-mode=places,round-precision=1]{      1841} & \num[scientific-notation=false,round-mode=places,round-precision=1]{    1536.3} & \num[scientific-notation=false,round-mode=places,round-precision=1]{      1666} \\
ps-2000-svd-50             & \num[fixed-exponent=10]{3328600e+4} & \num[fixed-exponent=10]{3330530e+4} & \num[fixed-exponent=10]{3329520e+4} & \num[fixed-exponent=10]{3329470e+4} & \num[scientific-notation=false,round-mode=places,round-precision=1]{       945} & \num[scientific-notation=false,round-mode=places,round-precision=1]{      2235} & \num[scientific-notation=false,round-mode=places,round-precision=1]{    1452.7} & \num[scientific-notation=false,round-mode=places,round-precision=1]{      1347} \\
\end{longtable}
}

{%
\scriptsize%
\begin{longtable}{l@{\hskip 4\tabcolsep}r@{\hskip 4\tabcolsep}r@{\hskip 4\tabcolsep}r@{\hskip 4\tabcolsep}r@{\hskip 8\tabcolsep}r@{\hskip 4\tabcolsep}r@{\hskip 4\tabcolsep}r@{\hskip 4\tabcolsep}r}
\toprule
{Group} & \multicolumn{4}{@{}c@{\hskip 8\tabcolsep}}{Cost} & \multicolumn{4}{c}{Running time}\\
\cmidrule(lr{16.8pt}){2-5} %
\cmidrule(lr){6-9}
{} &
\multicolumn{1}{c@{\hskip 4\tabcolsep}}{min} & \multicolumn{1}{c@{\hskip 4\tabcolsep}}{max} & \multicolumn{1}{c@{\hskip 4\tabcolsep}}{average} & \multicolumn{1}{c@{\hskip 12\tabcolsep}}{median} &
min & max & \multicolumn{1}{c@{\hskip 4\tabcolsep}}{avg} & \multicolumn{1}{c}{med} \\
\midrule \endfirsthead
{} &
\multicolumn{1}{c@{\hskip 4\tabcolsep}}{min} & \multicolumn{1}{c@{\hskip 4\tabcolsep}}{max} & \multicolumn{1}{c@{\hskip 4\tabcolsep}}{average} & \multicolumn{1}{c@{\hskip 12\tabcolsep}}{median} &
min & max & \multicolumn{1}{c@{\hskip 4\tabcolsep}}{avg} & \multicolumn{1}{c}{med} \\
\midrule \endhead
\bottomrule
\caption{Piecy on \texttt{LowerBound} instances with $n=10000$ and a piece size of $2000$, experiments belong to class II.\label{piecy-lowerbound}}\\%
\endfoot
k-10-d-10010-k-10-svd-15                    & \num[fixed-exponent=7]{  57174400} & \num[fixed-exponent=7]{  61050700} & \num[fixed-exponent=7]{58611740.0} & \num[fixed-exponent=7]{  58408400} & \num[scientific-notation=false,round-mode=places,round-precision=1]{        53} & \num[scientific-notation=false,round-mode=places,round-precision=1]{        76} & \num[scientific-notation=false,round-mode=places,round-precision=1]{      62.9} & \num[scientific-notation=false,round-mode=places,round-precision=1]{        58} \\
k-10-d-10010-k-10-svd-20                    & \num[fixed-exponent=7]{  54856900} & \num[fixed-exponent=7]{  56818600} & \num[fixed-exponent=7]{55788060.0} & \num[fixed-exponent=7]{  55809800} & \num[scientific-notation=false,round-mode=places,round-precision=1]{        56} & \num[scientific-notation=false,round-mode=places,round-precision=1]{        94} & \num[scientific-notation=false,round-mode=places,round-precision=1]{      69.1} & \num[scientific-notation=false,round-mode=places,round-precision=1]{        61} \\
k-50-d-10050-k-50-svd-75                    & \num[fixed-exponent=7]{  56855300} & \num[fixed-exponent=7]{  57715600} & \num[fixed-exponent=7]{57334940.0} & \num[fixed-exponent=7]{  57333700} & \num[scientific-notation=false,round-mode=places,round-precision=1]{        78} & \num[scientific-notation=false,round-mode=places,round-precision=1]{        78} & \num[scientific-notation=false,round-mode=places,round-precision=1]{      78.2} & \num[scientific-notation=false,round-mode=places,round-precision=1]{        78} \\
\end{longtable}
}

\begin{figure}
\includegraphics[]{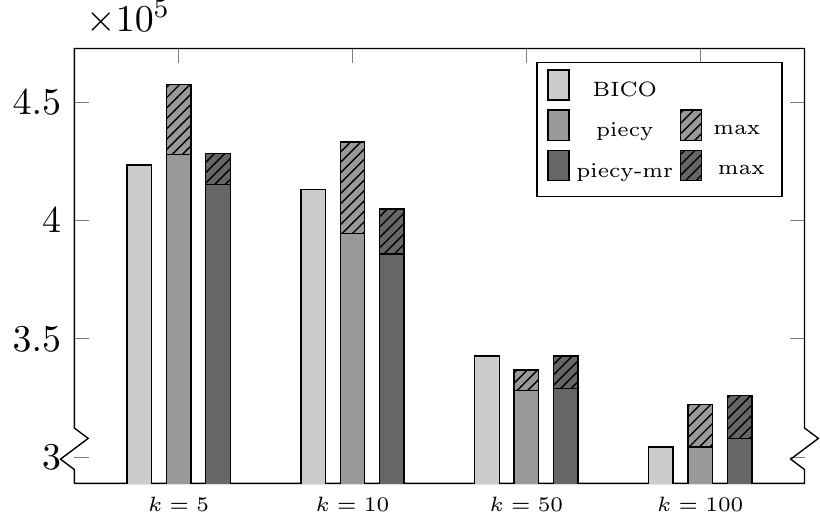}
\includegraphics[]{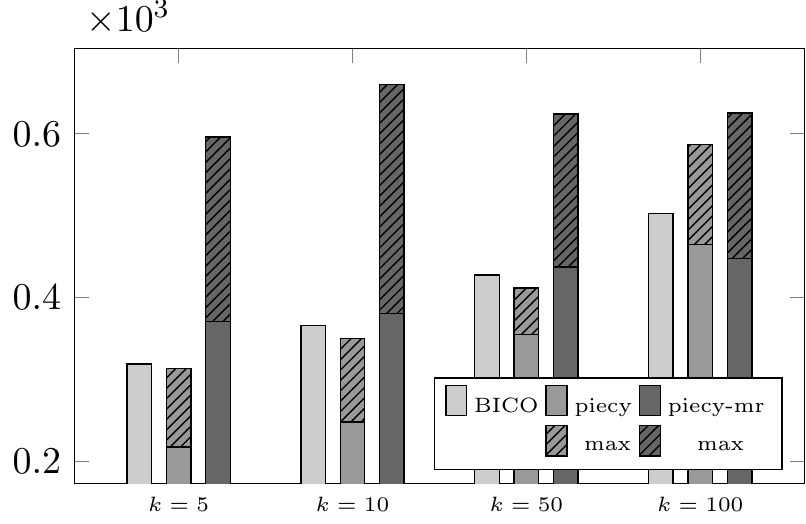}
\caption{Results for the \texttt{Caltech128} data set. Left side reports quality, right side run times. Variances stem from different parameters.\label{caltechdiagram}}
\end{figure}

\begin{figure}
\includegraphics[]{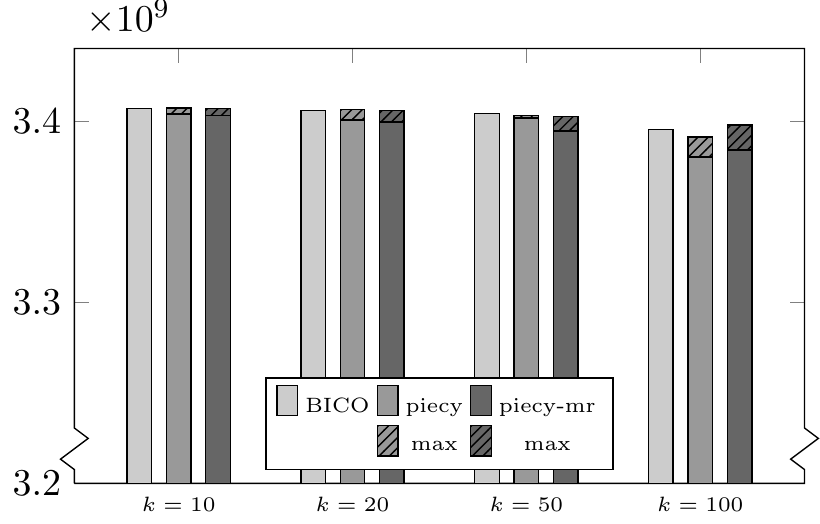}
\hfill
\includegraphics[]{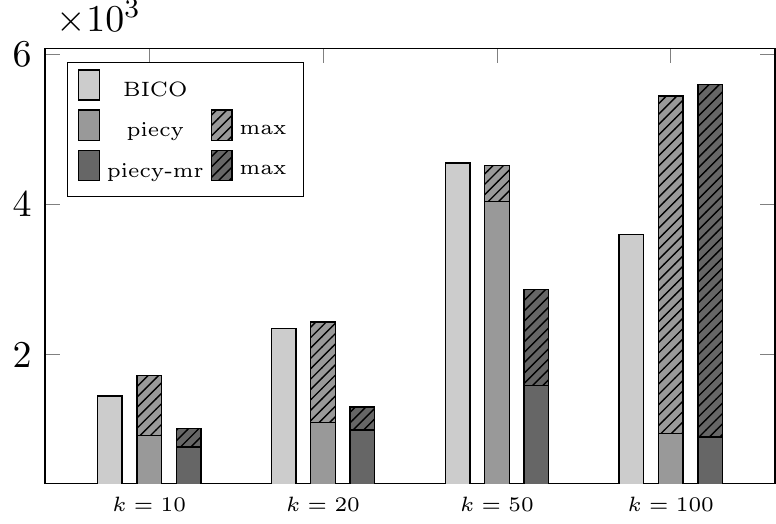}
\caption{Results for a \texttt{StructuredWithNoise} data set with $10^6$ points in $10^3$ dimensions. Left side reports quality, right side run times. Variances stem from different parameters.\label{swndiagram}}
\end{figure}

\subsubsection*{Piecy-mr.}

Piecy-mr also uses \emph{ps}, the piece size, as a parameter, as well as \emph{svd}, the number of dimensions to project to. The additional parameter \emph{np} is the number of pieces that are processed into the same BICO instance.

For \texttt{CalTech128}, the overhead of piecy-mr does not pay off and it performs worse than piecy. Results for this data set ar shown in Figure~\ref{caltechdiagram}
On the \texttt{LowerBound} test cases, piecy-mr is always slightly faster than BICO and comparable to piecy.
On the \texttt{Random} instances, piecy-mr is much faster than BICO, close to a factor of 2 on most test cases. This is in particular a much better running time than for piecy. The fact that \texttt{Random} has both a huge number of points and a high dimension means that the strength of piecy-mr shows and is not dominated by the overhead of the computation tree. The study of the three \texttt{StructuredWithNoise} data sets confirms this behaviour. In all three cases, the running time of piecy-mr is much faster or at least comparable to BICO with very few exceptions. This effect is particularly clear for the largest data set with one million points and a dimension of 1000, showing the speed of piecy-mr for large high-dimensional data sets. Figure~\ref{swndiagram} shows results for this data set. Notice that the large variance for piecy and piecy-mr is due to very different parameter choices. The best parameter choices yield a significant speed-up, particularly for large values of $k$.

{%
\scriptsize%
\begin{longtable}{l@{\hskip 4\tabcolsep}r@{\hskip 4\tabcolsep}r@{\hskip 4\tabcolsep}r@{\hskip 4\tabcolsep}r@{\hskip 8\tabcolsep}r@{\hskip 4\tabcolsep}r@{\hskip 4\tabcolsep}r@{\hskip 4\tabcolsep}r}
\toprule
{Group} & \multicolumn{4}{@{}c@{\hskip 8\tabcolsep}}{Cost} & \multicolumn{4}{c}{Running time}\\
\cmidrule(lr{16.8pt}){2-5} %
\cmidrule(lr){6-9}
{} &
\multicolumn{1}{c@{\hskip 4\tabcolsep}}{min} & \multicolumn{1}{c@{\hskip 4\tabcolsep}}{max} & \multicolumn{1}{c@{\hskip 4\tabcolsep}}{average} & \multicolumn{1}{c@{\hskip 12\tabcolsep}}{median} &
min & max & \multicolumn{1}{c@{\hskip 4\tabcolsep}}{avg} & \multicolumn{1}{c}{med} \\
\midrule \endfirsthead
{} &
\multicolumn{1}{c@{\hskip 4\tabcolsep}}{min} & \multicolumn{1}{c@{\hskip 4\tabcolsep}}{max} & \multicolumn{1}{c@{\hskip 4\tabcolsep}}{average} & \multicolumn{1}{c@{\hskip 12\tabcolsep}}{median} &
min & max & \multicolumn{1}{c@{\hskip 4\tabcolsep}}{avg} & \multicolumn{1}{c}{med} \\
\midrule \endhead
\bottomrule
\caption{Piecy-mr on \texttt{Caltech128}, experiments belong to class I.}\\%
\endfoot
\bottomrule
\caption[]{Piecy-mr on \texttt{Caltech128} (continued), experiments belong to class I.\label{piecy-mr-caltech}}\\%
\endlastfoot
{$k=5$}\\
\midrule
ps-5000-np-5-s-50                            & \num[fixed-exponent = 11]{42754800e+4} & \num[fixed-exponent = 11]{42911500e+4} & \num[fixed-exponent = 11]{42830920e+4} & \num[fixed-exponent = 11]{42824200e+4} & \num[scientific-notation=false,round-mode=places,round-precision=1]{       469} & \num[scientific-notation=false,round-mode=places,round-precision=1]{       504} & \num[scientific-notation=false,round-mode=places,round-precision=1]{     483.6} & \num[scientific-notation=false,round-mode=places,round-precision=1]{       479} \\
ps-5000-np-5-s-75                            & \num[fixed-exponent = 11]{42486500e+4} & \num[fixed-exponent = 11]{42864500e+4} & \num[fixed-exponent = 11]{42691880e+4} & \num[fixed-exponent = 11]{42654800e+4} & \num[scientific-notation=false,round-mode=places,round-precision=1]{       491} & \num[scientific-notation=false,round-mode=places,round-precision=1]{       514} & \num[scientific-notation=false,round-mode=places,round-precision=1]{     501.5} & \num[scientific-notation=false,round-mode=places,round-precision=1]{       500} \\
ps-5000-np-5-s-100                           & \num[fixed-exponent = 11]{42299700e+4} & \num[fixed-exponent = 11]{42925400e+4} & \num[fixed-exponent = 11]{42689720e+4} & \num[fixed-exponent = 11]{42729000e+4} & \num[scientific-notation=false,round-mode=places,round-precision=1]{       502} & \num[scientific-notation=false,round-mode=places,round-precision=1]{       548} & \num[scientific-notation=false,round-mode=places,round-precision=1]{     525.5} & \num[scientific-notation=false,round-mode=places,round-precision=1]{       525} \\
\end{longtable}
}

{%
\scriptsize %
\begin{longtable}{l@{\hskip 4\tabcolsep}r@{\hskip 4\tabcolsep}r@{\hskip 4\tabcolsep}r@{\hskip 4\tabcolsep}r@{\hskip 8\tabcolsep}r@{\hskip 4\tabcolsep}r@{\hskip 4\tabcolsep}r@{\hskip 4\tabcolsep}r}
\toprule
{Group} & \multicolumn{4}{@{}c@{\hskip 8\tabcolsep}}{Cost} & \multicolumn{4}{c}{Running time}\\
\cmidrule(lr{16.8pt}){2-5} %
\cmidrule(lr){6-9}
{} &
\multicolumn{1}{c@{\hskip 4\tabcolsep}}{min} & \multicolumn{1}{c@{\hskip 4\tabcolsep}}{max} & \multicolumn{1}{c@{\hskip 4\tabcolsep}}{average} & \multicolumn{1}{c@{\hskip 12\tabcolsep}}{median} &
min & max & \multicolumn{1}{c@{\hskip 4\tabcolsep}}{avg} & \multicolumn{1}{c}{med} \\
\midrule \endfirsthead
{} &
\multicolumn{1}{c@{\hskip 4\tabcolsep}}{min} & \multicolumn{1}{c@{\hskip 4\tabcolsep}}{max} & \multicolumn{1}{c@{\hskip 4\tabcolsep}}{average} & \multicolumn{1}{c@{\hskip 12\tabcolsep}}{median} &
min & max & \multicolumn{1}{c@{\hskip 4\tabcolsep}}{avg} & \multicolumn{1}{c}{med} \\
\midrule \endhead
\bottomrule
\caption{Piecy on \texttt{StructuredWithNoise}, experiments belong to class I.\label{piecy-SWN}}\\%
\endfoot
\bottomrule
\caption[]{Piecy on \texttt{StructuredWithNoise} (continued), experiments belong to class I.}\\%
\endlastfoot
{$k=5$}\\
\midrule
n-5000-d-1000 \\
\midrule
k-10-ps-2000-svd-10 & \num[fixed-exponent = 9]{170087e+4} & \num[fixed-exponent = 9]{170265e+4} & \num[fixed-exponent = 9]{170183e+4} & \num[fixed-exponent = 9]{170196e+4} & \num[scientific-notation=false,round-mode=places,round-precision=1]{       391} & \num[scientific-notation=false,round-mode=places,round-precision=1]{      1018} & \num[scientific-notation=false,round-mode=places,round-precision=1]{     703.4} & \num[scientific-notation=false,round-mode=places,round-precision=1]{       644} \\
k-10-ps-2000-svd-20 & \num[fixed-exponent = 9]{170063e+4} & \num[fixed-exponent = 9]{170408e+4} & \num[fixed-exponent = 9]{170246e+4} & \num[fixed-exponent = 9]{170256e+4} & \num[scientific-notation=false,round-mode=places,round-precision=1]{       442} & \num[scientific-notation=false,round-mode=places,round-precision=1]{      1155} & \num[scientific-notation=false,round-mode=places,round-precision=1]{     767.6} & \num[scientific-notation=false,round-mode=places,round-precision=1]{       736} \\
k-10-ps-2000-svd-50 & \num[fixed-exponent = 9]{170259e+4} & \num[fixed-exponent = 9]{170408e+4} & \num[fixed-exponent = 9]{170338e+4} & \num[fixed-exponent = 9]{170333e+4} & \num[scientific-notation=false,round-mode=places,round-precision=1]{       348} & \num[scientific-notation=false,round-mode=places,round-precision=1]{      1018} & \num[scientific-notation=false,round-mode=places,round-precision=1]{     554.4} & \num[scientific-notation=false,round-mode=places,round-precision=1]{       489} \\
\end{longtable}
}

{%
\scriptsize
\begin{longtable}{l@{\hskip 4\tabcolsep}r@{\hskip 4\tabcolsep}r@{\hskip 4\tabcolsep}r@{\hskip 4\tabcolsep}r@{\hskip 8\tabcolsep}r@{\hskip 4\tabcolsep}r@{\hskip 4\tabcolsep}r@{\hskip 4\tabcolsep}r}
\toprule
{Group} & \multicolumn{4}{@{}c@{\hskip 8\tabcolsep}}{Cost} & \multicolumn{4}{c}{Running time}\\
\cmidrule(lr{16.8pt}){2-5} %
\cmidrule(lr){6-9}
{} &
\multicolumn{1}{c@{\hskip 4\tabcolsep}}{min} & \multicolumn{1}{c@{\hskip 4\tabcolsep}}{max} & \multicolumn{1}{c@{\hskip 4\tabcolsep}}{average} & \multicolumn{1}{c@{\hskip 12\tabcolsep}}{median} &
min & max & \multicolumn{1}{c@{\hskip 4\tabcolsep}}{avg} & \multicolumn{1}{c}{med} \\
\midrule \endfirsthead
{} &
\multicolumn{1}{c@{\hskip 4\tabcolsep}}{min} & \multicolumn{1}{c@{\hskip 4\tabcolsep}}{max} & \multicolumn{1}{c@{\hskip 4\tabcolsep}}{average} & \multicolumn{1}{c@{\hskip 12\tabcolsep}}{median} &
min & max & \multicolumn{1}{c@{\hskip 4\tabcolsep}}{avg} & \multicolumn{1}{c}{med} \\
\midrule \endhead
\bottomrule
\caption{Piecy-mr on \texttt{random} instances with $d=1000$ and $n=10^6$, experiments belong to class II.}\\%
\endfoot
\bottomrule
\caption[]{Piecy-mr on \texttt{random} instances with $d=1000$ and $n=10^6$ (continued), experiments belong to class II.}\\%
\endlastfoot
\multicolumn{2}{l}{$k=10$, piece size $2000$}\\
\midrule
np-10-svd-10          & \num[fixed-exponent=10]{3325760e+4} & \num[fixed-exponent=10]{3327860e+4} & \num[fixed-exponent=10]{3326928e+4} & \num[fixed-exponent=10]{3327140e+4} & \num[scientific-notation=false,round-mode=places,round-precision=1]{       768} & \num[scientific-notation=false,round-mode=places,round-precision=1]{       863} & \num[scientific-notation=false,round-mode=places,round-precision=1]{     817.4} & \num[scientific-notation=false,round-mode=places,round-precision=1]{       820} \\
np-10-svd-20          & \num[fixed-exponent=10]{3327570e+4} & \num[fixed-exponent=10]{3330180e+4} & \num[fixed-exponent=10]{3328772e+4} & \num[fixed-exponent=10]{3328810e+4} & \num[scientific-notation=false,round-mode=places,round-precision=1]{       902} & \num[scientific-notation=false,round-mode=places,round-precision=1]{       953} & \num[scientific-notation=false,round-mode=places,round-precision=1]{     917.4} & \num[scientific-notation=false,round-mode=places,round-precision=1]{       906} \\
np-10-svd-50          & \num[fixed-exponent=10]{3328340e+4} & \num[fixed-exponent=10]{3330190e+4} & \num[fixed-exponent=10]{3329268e+4} & \num[fixed-exponent=10]{3329070e+4} & \num[scientific-notation=false,round-mode=places,round-precision=1]{       873} & \num[scientific-notation=false,round-mode=places,round-precision=1]{       925} & \num[scientific-notation=false,round-mode=places,round-precision=1]{     902.1} & \num[scientific-notation=false,round-mode=places,round-precision=1]{       906} \\
\end{longtable}
}

{%
\scriptsize
\begin{longtable}{l@{\hskip 4\tabcolsep}r@{\hskip 4\tabcolsep}r@{\hskip 4\tabcolsep}r@{\hskip 4\tabcolsep}r@{\hskip 8\tabcolsep}r@{\hskip 4\tabcolsep}r@{\hskip 4\tabcolsep}r@{\hskip 4\tabcolsep}r}
\toprule
{Group} & \multicolumn{4}{@{}c@{\hskip 8\tabcolsep}}{Cost} & \multicolumn{4}{c}{Running time}\\
\cmidrule(lr{16.8pt}){2-5} %
\cmidrule(lr){6-9}
{} &
\multicolumn{1}{c@{\hskip 4\tabcolsep}}{min} & \multicolumn{1}{c@{\hskip 4\tabcolsep}}{max} & \multicolumn{1}{c@{\hskip 4\tabcolsep}}{average} & \multicolumn{1}{c@{\hskip 12\tabcolsep}}{median} &
min & max & \multicolumn{1}{c@{\hskip 4\tabcolsep}}{avg} & \multicolumn{1}{c}{med} \\
\midrule \endfirsthead
{} &
\multicolumn{1}{c@{\hskip 4\tabcolsep}}{min} & \multicolumn{1}{c@{\hskip 4\tabcolsep}}{max} & \multicolumn{1}{c@{\hskip 4\tabcolsep}}{average} & \multicolumn{1}{c@{\hskip 12\tabcolsep}}{median} &
min & max & \multicolumn{1}{c@{\hskip 4\tabcolsep}}{avg} & \multicolumn{1}{c}{med} \\
\midrule \endhead
\bottomrule
\caption{Piecy-mr on \texttt{LowerBound} instances with $d=10010$, $n=10000$ and $m,k=10$. The piece size is fixed to $2000$, the number of pieces is fixed to $10$. Experiments belong to class II.}\\%
\endfoot
Group                                                                  &   Min cost &   Max cost &   Avg cost & Median cost &   Min time &   Max time &   Avg time & Median time \\
\midrule
svd-15                 & \num[fixed-exponent=7]{  59050100} & \num[fixed-exponent=7]{  59050100} & \num[fixed-exponent=7]{59050100.0} & \num[fixed-exponent=7]{  59050100} & \num[scientific-notation=false,round-mode=places,round-precision=1]{        58} & \num[scientific-notation=false,round-mode=places,round-precision=1]{        68} & \num[scientific-notation=false,round-mode=places,round-precision=1]{      62.9} & \num[scientific-notation=false,round-mode=places,round-precision=1]{        63} \\
svd-20                 & \num[fixed-exponent=7]{  56163800} & \num[fixed-exponent=7]{  56163800} & \num[fixed-exponent=7]{56163800.0} & \num[fixed-exponent=7]{  56163800} & \num[scientific-notation=false,round-mode=places,round-precision=1]{        62} & \num[scientific-notation=false,round-mode=places,round-precision=1]{        89} & \num[scientific-notation=false,round-mode=places,round-precision=1]{      70.0} & \num[scientific-notation=false,round-mode=places,round-precision=1]{        66} \\
\end{longtable}
}

{%
\scriptsize
\begin{longtable}{l@{\hskip 4\tabcolsep}r@{\hskip 4\tabcolsep}r@{\hskip 4\tabcolsep}r@{\hskip 4\tabcolsep}r@{\hskip 8\tabcolsep}r@{\hskip 4\tabcolsep}r@{\hskip 4\tabcolsep}r@{\hskip 4\tabcolsep}r}
\toprule
{Group} & \multicolumn{4}{@{}c@{\hskip 8\tabcolsep}}{Cost} & \multicolumn{4}{c}{Running time}\\
\cmidrule(lr{16.8pt}){2-5} %
\cmidrule(lr){6-9}
{} &
\multicolumn{1}{c@{\hskip 4\tabcolsep}}{min} & \multicolumn{1}{c@{\hskip 4\tabcolsep}}{max} & \multicolumn{1}{c@{\hskip 4\tabcolsep}}{average} & \multicolumn{1}{c@{\hskip 12\tabcolsep}}{median} &
min & max & \multicolumn{1}{c@{\hskip 4\tabcolsep}}{avg} & \multicolumn{1}{c}{med} \\
\midrule \endfirsthead
{} &
\multicolumn{1}{c@{\hskip 4\tabcolsep}}{min} & \multicolumn{1}{c@{\hskip 4\tabcolsep}}{max} & \multicolumn{1}{c@{\hskip 4\tabcolsep}}{average} & \multicolumn{1}{c@{\hskip 12\tabcolsep}}{median} &
min & max & \multicolumn{1}{c@{\hskip 4\tabcolsep}}{avg} & \multicolumn{1}{c}{med} \\
\midrule \endhead
\bottomrule
\caption{Piecy-MR on \texttt{StructuredWithNoise} with $n=1000000$, $d=500$ and $k=50$. Experiments belong to class III.}\\%
\endfoot
ps-10000-np-5-svd-75          & \num[fixed-exponent=9]{333530e+4} & \num[fixed-exponent=9]{334804e+4} & \num[fixed-exponent=9]{334092e+4} & \num[fixed-exponent=9]{334067e+4} & \num[scientific-notation=false,round-mode=places,round-precision=1]{       663} & \num[scientific-notation=false,round-mode=places,round-precision=1]{      1086} & \num[scientific-notation=false,round-mode=places,round-precision=1]{     843.8} & \num[scientific-notation=false,round-mode=places,round-precision=1]{       829} \\
ps-10000-np-10-svd-75         & \num[fixed-exponent=9]{333618e+4} & \num[fixed-exponent=9]{335380e+4} & \num[fixed-exponent=9]{334390e+4} & \num[fixed-exponent=9]{334198e+4} & \num[scientific-notation=false,round-mode=places,round-precision=1]{       453} & \num[scientific-notation=false,round-mode=places,round-precision=1]{       769} & \num[scientific-notation=false,round-mode=places,round-precision=1]{     632.2} & \num[scientific-notation=false,round-mode=places,round-precision=1]{       629} \\
ps-10000-np-10-svd-50         & \num[fixed-exponent=9]{333499e+4} & \num[fixed-exponent=9]{334482e+4} & \num[fixed-exponent=9]{333826e+4} & \num[fixed-exponent=9]{333751e+4} & \num[scientific-notation=false,round-mode=places,round-precision=1]{       426} & \num[scientific-notation=false,round-mode=places,round-precision=1]{       873} & \num[scientific-notation=false,round-mode=places,round-precision=1]{     627.0} & \num[scientific-notation=false,round-mode=places,round-precision=1]{       609} \\
\end{longtable}
}

{%
\scriptsize
\begin{longtable}{l@{\hskip 4\tabcolsep}r@{\hskip 4\tabcolsep}r@{\hskip 4\tabcolsep}r@{\hskip 4\tabcolsep}r@{\hskip 8\tabcolsep}r@{\hskip 4\tabcolsep}r@{\hskip 4\tabcolsep}r@{\hskip 4\tabcolsep}r}
\toprule
{Group} & \multicolumn{4}{@{}c@{\hskip 8\tabcolsep}}{Cost} & \multicolumn{4}{c}{Running time}\\
\cmidrule(lr{16.8pt}){2-5} %
\cmidrule(lr){6-9}
{} &
\multicolumn{1}{c@{\hskip 4\tabcolsep}}{min} & \multicolumn{1}{c@{\hskip 4\tabcolsep}}{max} & \multicolumn{1}{c@{\hskip 4\tabcolsep}}{average} & \multicolumn{1}{c@{\hskip 12\tabcolsep}}{median} &
min & max & \multicolumn{1}{c@{\hskip 4\tabcolsep}}{avg} & \multicolumn{1}{c}{med} \\
\midrule \endfirsthead
{} &
\multicolumn{1}{c@{\hskip 4\tabcolsep}}{min} & \multicolumn{1}{c@{\hskip 4\tabcolsep}}{max} & \multicolumn{1}{c@{\hskip 4\tabcolsep}}{average} & \multicolumn{1}{c@{\hskip 12\tabcolsep}}{median} &
min & max & \multicolumn{1}{c@{\hskip 4\tabcolsep}}{avg} & \multicolumn{1}{c}{med} \\
\midrule \endhead
\bottomrule
\caption{Piecy-MR on \texttt{StructuredWithNoise}. Experiments belong to class II.}\\%
\endfoot
n-5000-d-1000 \\
\midrule
k-10-ps-2000-np-10-svd-10 & \num[fixed-exponent = 9]{170159e+4} & \num[fixed-exponent = 9]{170289e+4} & \num[fixed-exponent = 9]{170226e+4} & \num[fixed-exponent = 9]{170233e+4} & \num[scientific-notation=false,round-mode=places,round-precision=1]{       346} & \num[scientific-notation=false,round-mode=places,round-precision=1]{       397} & \num[scientific-notation=false,round-mode=places,round-precision=1]{     375.9} & \num[scientific-notation=false,round-mode=places,round-precision=1]{       382} \\
k-10-ps-2000-np-10-svd-20 & \num[fixed-exponent = 9]{170147e+4} & \num[fixed-exponent = 9]{170345e+4} & \num[fixed-exponent = 9]{170200e+4} & \num[fixed-exponent = 9]{170205e+4} & \num[scientific-notation=false,round-mode=places,round-precision=1]{       401} & \num[scientific-notation=false,round-mode=places,round-precision=1]{       456} & \num[scientific-notation=false,round-mode=places,round-precision=1]{     422.4} & \num[scientific-notation=false,round-mode=places,round-precision=1]{       422} \\
k-10-ps-2000-np-10-svd-50 & \num[fixed-exponent = 9]{170251e+4} & \num[fixed-exponent = 9]{170383e+4} & \num[fixed-exponent = 9]{170333e+4} & \num[fixed-exponent = 9]{170365e+4} & \num[scientific-notation=false,round-mode=places,round-precision=1]{       410} & \num[scientific-notation=false,round-mode=places,round-precision=1]{       482} & \num[scientific-notation=false,round-mode=places,round-precision=1]{     434.3} & \num[scientific-notation=false,round-mode=places,round-precision=1]{       430} \\
\end{longtable}
}

\subsubsection*{Conclusion.} The experiments show the potential speed-up by using piecy and piecy-mr. When choosing the algorithm, one should take the dimensions of the input matrix into account. For large dimension but a moderate number of points, piecy is ideal since it reduces the dimension effectively with little overhead. For data sets where the dimension is high and the number of points is also high, the additional overhead of piecy-mr pays off.

\subsubsection*{Acknowledgements.}
We thank Cameron Musco and Chris Schwiegelshohn for insightful discussions on the topic of this paper, Hendrik Fichtenberger and Lukas Pradel for sharing some pieces of source code and Jan Stallmann and René Grzeszick for providing the \texttt{LowerBound} and \texttt{CalTech} data sets.

\newpage

\bibliographystyle{plain}
\enlargethispage{\baselineskip}
\bibliography{references}

\end{document}